\documentclass[]{aa}
\usepackage{graphicx}
\usepackage[normalem]{ulem} 
\usepackage{txfonts}
\usepackage{amssymb}
\usepackage{enumerate}
\usepackage{subfigure}
\usepackage[colorlinks=true,allcolors=blue]{hyperref}%
\newcommand{\mycomment}[1]{}

\renewcommand{\vec}[1]{{\mathbf #1}}

\newcommand{\llnr}[1]{{\bf \color{magenta}{[]}} \color{black}} 

\begin{document}

 \title{Self-consistent propagation of flux ropes in realistic coronal simulations}

 \author{L. Linan\inst{1} \and F. Regnault\inst{2} \and B. Perri\inst{3} \and M. Brchnelova\inst{1} \and B. Kuzma\inst{4} \and A. Lani\inst{5} \and S. Poedts\inst{1,6} \and B. Schmieder\inst{1,7,8} }

 \institute{Centre for Mathematical Plasma-Astrophysics, Department of Mathematics, KU Leuven, Celestijnenlaan 200B, 3001 Leuven, Belgium \\
		 \and Space Science Center, Institute for the Study of Earth, Oceans, and Space, and Department of Physics and Astronomy, University of New Hampshire, College Road, Durham, NH 03824, United States \\
  \and Université Paris-Saclay, Université Paris Cité, CEA, CNRS, AIM, 91191, Gif-sur-Yvette, France \\
  \and Institute of Space Science and Applied Technology, Harbin Institute of Technology, Shenzhen 518055, People's Republic of China \\ 
  \and Von Karman Institute For Fluid Dynamics, Waterloosesteenweg 72. B-1640, Sint-Genesius-Rode, Brussels, Belgium \\
  \and Institute of Physics, University of Maria Curie-Sk\l odowska, ul.\ Radziszewskiego 10, 20-031 Lublin, Poland \\
  \and LESIA, Observatoire de Paris, Université PSL, CNRS, Sorbonne Université, Université de Paris, 5 place Jules Janssen, 92190 Meudon, France\\
  \and University of Glasgow, School of Physics and Astronomy, Glasgow, G128QQ, Scotland
  }

 \date{Received XXXX; accepted XXXX}

 
 \abstract
 {In order to anticipate the geoeffectiveness of coronal mass ejections (CMEs), heliospheric simulations are used to propagate transient structures injected at 0.1 astronomical units (AU). Without direct measurements near the Sun, the properties of these injected CMEs must be derived from models coming from observations or numerical simulations and thus they contain a lot of uncertainty.}
 {The aim of this paper is to demonstrate the possible use of the new coronal model COCONUT to compute a detailed representation of a numerical CME at 0.1~AU, after its injection at the solar surface and propagation in a realistic solar wind, as derived from observed magnetograms.}
 {We present the implementation and propagation of modified Titov-D\'emoulin (TDm) flux ropes in the COCONUT 3D MHD coronal model. The background solar wind is reconstructed in order to model two opposite configurations representing a solar activity maximum and minimum respectively. Both were derived from magnetograms which were obtained by the Helioseismic and Magnetic Imager (HMI) onboard the Solar Dynamic Observatory (SDO) satellite. We track the propagation of 24 flux ropes, which differ only by their initial magnetic flux. We especially investigate the geometry of the flux rope during the early stages of the propagation as well as the influence of its initial parameters and solar wind configuration on 1D profiles derived at 0.1~AU.}
 {At the beginning of the propagation, the shape of the flux ropes varies between simulations during low and high solar activity. We find dynamics that are consistent with the standard CME model, such as the pinching of the legs and the appearance of post-flare loops. Despite the differences in geometry, the synthetic density and magnetic field time profiles at 0.1~AU are very similar in both solar wind configurations. These profiles are similar to those observed further in the heliosphere and suggest the presence of a magnetic ejecta composed of the initially implemented flux rope and a sheath ahead of it. Finally, we uncover relationships between the properties of the magnetic ejecta, such as density or speed and the initial magnetic flux of our flux ropes.}
 {The implementation of the modified Titov-D\'emoulin flux rope in COCONUT enables us to retrieve the major properties of CMEs at 0.1~AU for any phase of the solar cycle. When combined with heliospheric simulations, COCONUT could lead to more realistic and self-consistent CME evolution models, and thus more reliable predictions.} 


 \maketitle

\section{Introduction} \label{sec:Introduction}

Solar flares are sudden bursts of magnetic energy which are released from localized regions on the solar atmosphere called active regions (ARs). These emissions of radiation can be accompanied by the expulsion of magnetic plasma from the corona into the interplanetary medium. This is known as a Coronal Mass Ejection (CME) and observed with white-light coronagraphs. Solar activity follows an 11-year cycle, with a minimum of activity where only few CMEs observed and a maximum of activity characterized by increased magnetic activity levels \citep{Hathaway15}.

The solar activity can have various impacts on the Earth's magnetic environment, and thus on human technologies \citep{Bothmer07}. As a result, many industrial sectors (e.g. energy, telecommunication, transportation) may be negatively affected \citep{Hapgood10}. In addition to the damage that bursts of energetic particles can cause to space- and ground-based technologies, the aircrews may also be impacted by solar radiation \citep{Griffiths12}. To minimize risks and potential costs, it is necessary to have a thorough understanding of the physics of the Sun-Earth transients and the processes that cause space-weather events.

To improve the protection of both on-board crews and electronic devices, it is necessary to be able to predict the eruptivity of active regions sufficiently in advance \citep{Schrijver15}. There are various forecasting methods with varying levels of effectiveness \citep{Barnes16}, mainly using machine learning or deep learning artificial intelligence algorithms \citep[e.g.][]{Park18,Park20}. These methods often involve the parameterization of observed solar data in order to characterize the targeted active region \citep[e.g.][]{Bobra15,Bobra16,Falconer08,Falconer11,Li20}. From vector solar magnetograms various magnetic field parameters are calculated and used as eruptive signatures \citep{Guennou17}. The quantities at the heart of these algorithms rely upon magnetic field \citep[e.g.\ the free magnetic energy][]{Leka03}, current properties \citep[e.g.\ the current density][]{Zhang01,Torok14} or the magnetic field polarity inversion line properties \citep{Falconer08}. However, no algorithm is yet able to predict all solar activity, partly due to the rarity of major X-class events \citep{Aschwanden12}. In recent years, a new quantity defined as the ratio of the current carrying helicity to the relative helicity \citep{Pariat17} has been found to be a good indicator for the eruptivity of numerical magnetic configurations in 3D parametric simulations \citep{Linan18,Zuccarello18} and also for observed active regions \citep{James18,Moraitis19,Price19,Thalmann19,Thalmann21,Gupta21,Lumme22}. However, this criterion is not yet used in automatic forecasting models due to the complexity of computing it in the solar atmosphere \citep{Linan18,Linan20}.

Even if we were able to predict an eruptive flare before it occurs, we would also need to be able to track its propagation and evolution throughout the heliosphere, because not all CMEs impact Earth, and not all CMEs have the same properties and therefore the same geoeffectiveness \citep[cf.][for a compilation of CME profiles from 20 years of in-situ observations]{Regnault20}. Specifically, when the magnetic field lines of a CME and the Earth's magnetosphere have opposite orientations, magnetic reconnection can occur, resulting in geomagnetic storms with potentially significant impact on the Earth's environment \citep{Gonzalez94}.

In the interplanetary medium, CME signatures are probed at different distances from the Earth by many spacecraft such as ACE \citep{Chiu98}, Parker Solar Probe \citep{Fox16}, STEREO A and B \citep{Kaiser07}, Solar Orbiter \citep{Muller20} and WIND \citep{Harten95}. However, there are not enough satellites to perfectly determine the structure of a CME throughout its propagation \citep{Demoulin10}. To reconstruct efficiently the complex 3D structure of a CME from a limited set of 1D observations, one possibility is then to use multiple viewpoint reconstruction techniques \citep[e.g][]{Rodari18}. However, these techniques rely on approximations of the overall structure of the CME. It is worth noting that the future spatial mission PUNCH (Polarimeter to UNify the Corona and Heliosphere) will provide four new spacecrafts equipped with narrow and wide field imagers aimed at producing continuous 3D images of the solar corona \citep{DeForest22}. This should lead to a better understanding and determination of deformation effects that CMEs can undergo as they propagate such as expansion, aging and erosion \citep{Demoulin08}.

Observations can be supplemented with numerical simulations by various 3D magnetohydrodynamic (MHD) solvers, such as ENLIL \citep{Odstrcil03}, EUHFORIA \citep{Pomoell18}, MS-FLUKSS \citep{Singh18} and SUSANOO-CME \citep{Shiota16}, all capable of tracking CME propagation in a realistic description of the background solar wind. The latter is usually based on the empirical Wang-Sheeley-Arge (WSA) coronal model, where its properties are derived from the solar wind speed and the expansion of coronal magnetic field lines \citep{Wang90}. In these simulations, CMEs are inserted at 0.1~AU ($21.5\;R_{\odot}$) by modifying the solar wind properties (e.g.\ speed, density, magnetic field components). The parameters of the CMEs are derived from 3D models. Among them, we can mention the cone model describing the CME as an unmagnetized plasma with a self-similar expansion \citep{Xie04}, toroidal-like models where the CME is a flux rope connected to the Sun \citep[e.g.\ "Flux Rope in 3D", FRi3D][]{Isavnin16,Maharana22} and the linear force-free spheromak model representing a CME with a global spherical shape \citep{Chandrasekhar57,Verbeke19}. The initial geometric and magnetic parameters of these models can be partially derived from remote-sensing observations. In particular, the magnetic flux and helicity can be deduced from magnetograms while the geometric and kinematic parameters are obtained using the graduated cylindrical shell model based on white-light images \citep{Scolini19,Maharana22}. Finally, the accuracy of the forecast provided by the simulation depends on the accuracy of the background solar wind, the chosen CME models and their mutual interaction.

The main limitation of inserting the CME only at 0.1~AU is that it ignores all the physics that occurs lower in the solar corona. The CMEs do not evolve in the solar corona before its insertion and they are impossible to track in the currently used corona model (WSA) since it not a time dependent 3D simulation but only a set of empirical relations. Hence, the inserted CME does not interact with the solar wind before the inner boundary. This interaction is, however, crucial for accurately determining its geometric and magnetic characteristics \citep[cf.][for a description of the first stages of an eruption]{Green18}. For example, \citet{Asvestari22}, using the spheromak model, found that the CME tilt depends on ambient magnetic field strength and orientation. The interaction with the solar wind can also lead to a deflection of the CME into a streamer \citep{Zuccarello11}.

To address this limitation, this paper introduces the propagation of magnetic flux ropes in a recently implemented 3D MHD coronal plasma solver called COolfluid COroNal UnsTructured (COCONUT) \citep{Perri2022}. The computational MHD model in COCONUT was originally developed to replace the empirical model which is currently used by the EUropean Heliospheric FORecasting Information Asset (EUHFORIA) \citep{Poedts2020_euhforia}. The current version of COCONUT uses a polytropic heating, which means that the thermodynamics of the model is simplified while the magnetic configuration is realistic. In the polytropic model, we assume the ratio of specific heats, $\gamma$, to be set to 1.05 and neglect the coronal heating, radiation and conduction terms, whose inclusion is a focus of ongoing efforts. The main objective of this work is to demonstrate the potential use of COCONUT to track the propagation of flux ropes in the corona and to provide at $21.5\;R_{\odot}$ the necessary information for heliospheric simulations. The boundary conditions imposed on EUHFORIA (or other heliospheric models) should be more accurate than those provided by an independent CME model (e.g.\ spheromak) and therefore produce more realistic forecasts. To achieve the aforementioned objective, it was decided not to model a real CME event, which would have required matching the properties of the flux rope with its active region of origin; but instead, to follow the propagation of several "unobserved" flux ropes, which allows for changing the solar wind configuration as well as the properties of the flux rope to determine the advantages and limitations of COCONUT.

Recently, \citet{Regnault23} studied the propagation of a flux rope from the low corona up to 1~AU using the PLUTO multi-physics code \citep{Mignone2007}. However, in \citet{Regnault23}, the magnetic field configuration of the solar wind is dipolar while in this study the propagation is examined in two magnetic realistic configurations driven by magnetograms corresponding to a maximum and a minimum of solar activity. Additionally, COCONUT and PLUTO are two different coronal models with different grids, boundary conditions and numerical schemes. The COCONUT numerical parameters have been optimized to provide the best comparison with observations \citep{Brchnelova2022a, Brchnelova2022b}. Another difference is that COCONUT uses an implicit solver, while PLUTO uses an explicit solver, which means that the COCONUT code can run within operational times \citep{Perri2022}, making it more suitable for actual forecasting. 

The development of coronal MHD models has progressed significantly over the years with various approaches and techniques being employed to model the solar corona. The first models were based on idealized plasma conditions and did not incorporate observational data into their boundary conditions \citep[e.g][]{Pneuman71}. However, data-driven models have since been developed, such as the 3D MHD model of \citet{Linker99}, which was validated through white lights images of eclipses and interplanetary observations \citep[see also][]{Mikic96}.

Among the other existing 3D MHD models based on systematic comparisons to both in-situ and remote-sensing data for validation, some allow the study of physical processes that are not yet incorporated in COCONUT. Examples include the steady global corona model with turbulence transport and heating of \citet{Usmanov96,Usmanov14,Usmanov18,Chhiber21}, the Wind-Predict-AW code \citep{Reville2020, Reville2021, Parenti2022} or the Alfvén Wave Solar atmosphere Model \citep[AWSoM;][]{van10,Van15,Sachdeva19} that allow a realistic description of the temperature and density distribution thanks to the treatment of wave dissipation, heat conduction, and radiative cooling. This model is already implemented within the Space Weather Modeling Framework \citep[SWMF;][]{toth12}. The Magnetohydrodynamics Around a Sphere (MAS) solver developed by Predictive Science, Inc. is also one of the most advanced models and is based on time-dependent resistive MHD, including comprehensive energy transport mechanisms from radiation and thermal conduction to Alfvén wave heating \citep{Mikic99,Mikic18}. 

Although these different models possess apparent advantages, it is worth noting that some of them demand substantial numerical resources to generate a realistic 3D depiction of the solar corona. As a result, the COCONUT solver's implicit time scheme, combined with its highly scalable parallel architecture distinguishes it from other solvers and enhances its effectiveness in solving MHD problems quickly and accurately. This justifies our selection of COCONUT solver for this work.

The structure of our paper is as follows. We begin by introducing the COCONUT code that is used to create two realistic solar wind configurations based on HMI magnetograms from a quiet and an active Sun (cf.\ Sect.~\ref{sec:setupt}). Then, we present the flux rope model which has been used in this study (cf.\ Sect.~\ref{sec:TDM}). Special attention will be paid to how the flux rope is implemented in both two solar wind backgrounds (cf.\ Sect.~\ref{sec:implementation}) and to the initial parameters that differentiate the simulations. In section \ref{sec:Prop} we describe the early stages of propagation and the limitations imposed by high-speed streams, while in section \ref{sec:profiles} we focus on density, magnetic field and velocity profiles at $21.5$ solar radii. In particular, we discuss the impact of the solar wind and the magnetic flux of the flux ropes on these profiles.


\section{The numerical framework} \label{sec:setupt}
\subsection{COCONUT}\label{sec:Coconut}

\begin{figure*}[ht!]
\centering
\includegraphics[width=0.85\textwidth]{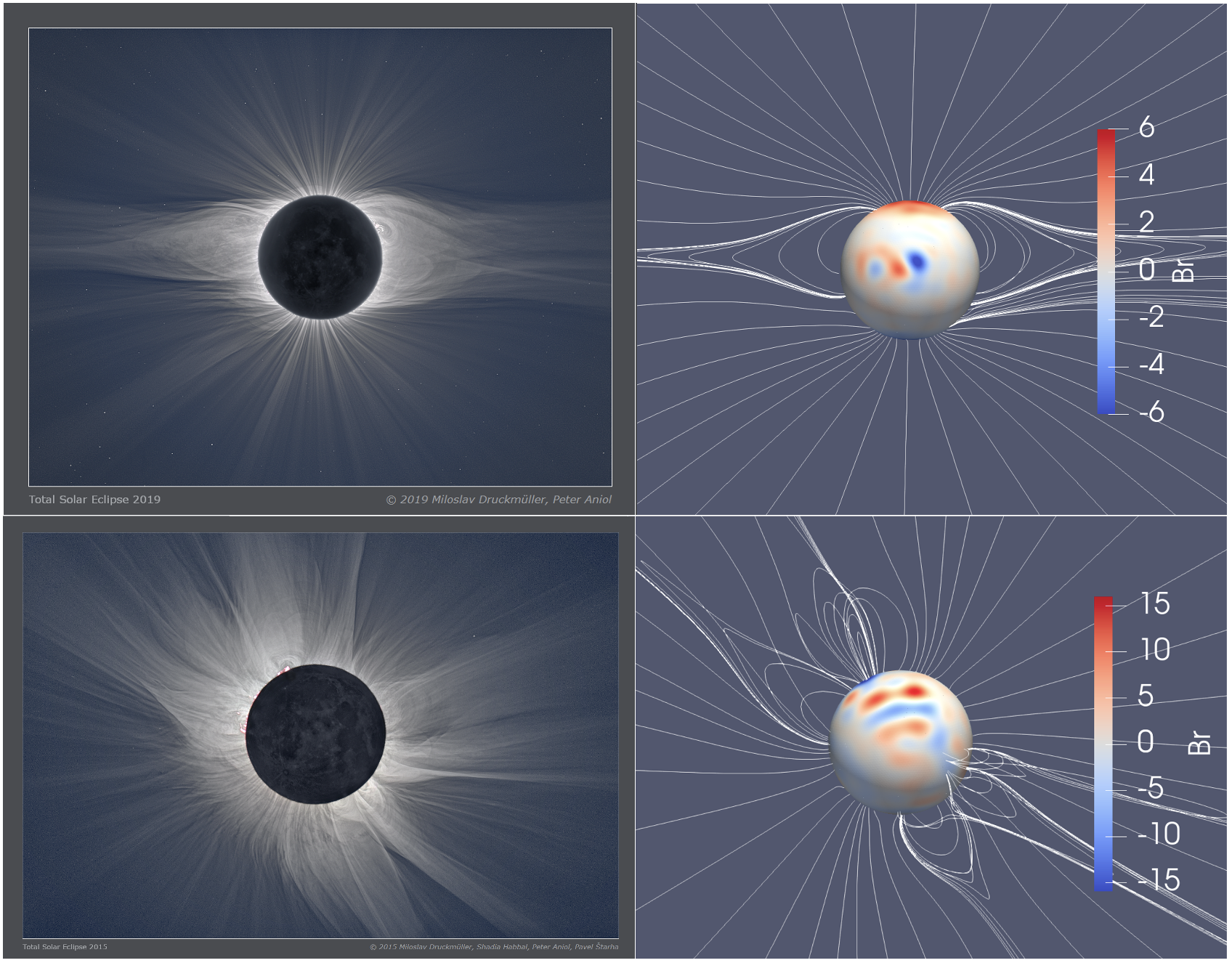}
\caption{Comparison of the solar eclipse images (left), and numerical results (right), for the case of minimum (top) and maximum (bottom) of the solar activity. 
The minimum activity image corresponds to solar eclipse of 2nd July 2019 (© 2019 Miloslav Druckm\"uller, Peter Aniol), the maximum activity image to solar eclipse of 20th March 2015 (© 2015 Miloslav Druckm\"uller, Shadia Habbal, Peter Aniol, Pavel Starha). The red-blue sphere colormap represents radial component of photospheric magnetic field expressed in Gauss, white lines represent magnetic field lines of the magnetic coronal structures in the plane of sight. The reconstructed main streamers align closely with the position and shape of coronal structures observed in white light.}
\label{fig:minmax}
\end{figure*}

In this study, we use the COCONUT code, which is based upon the Computational Object-Oriented Libraries for Fluid Dynamics (COOLFluiD) platform. COOLFluiD was originally developed for multi-physics and, particularly, flow/plasma simulations \citep{Lani2005, Kimpe2005, Lani2013}. The COCONUT solver has been extensively described, verified and validated in \citet{Perri2022} and \citet{Kuźma23}. The second-order finite volume (FV) scheme that COCONUT uses is implicit, which means that for steady-state simulations, Courant-Friedrichs-Lewy (CFL) numbers much higher than 1 can be used, enhancing the performance of the code considerably when compared to other state-of-art MHD coronal models. 

In addition, the grid is unstructured, which means that more advanced grid refinement techniques \citep[e.g.\ r-adaptative algorithms][]{BENAMEUR21} and flux reconstruction \citep[high-order Flux Reconstruction algorithms][]{Vandenhoeck2019} can be used in the future to further enhance the performance. The mesh used is a 6th-level subdivision of the geodesic polyhedron with 1.5 million cells. More details about the grid and its effects in COCONUT are presented in \citet{Brchnelova2022a}.

The code uses a second-order accurate FV discretization along with the Artificial Compressibility Analogy \citep{chorin1997} (similar to the hyperbolic divergence cleaning (HDC) of \citet{Dedner2002}) to ensure that the divergence of the $\mathbf{B}$-field remains close to zero. The following conservative formulation of the MHD equations is solved by COCONUT \citep{Perri2022}:

\begin{equation}
\frac{\partial}{\partial t}\left(\begin{array}{c}
\rho \\
\rho \vec{v} \\
\vec{B} \\
E \\
\phi \\
\end{array}\right)+\vec{\nabla} \cdot\left(\begin{array}{c}
\rho \vec{v} \\
\rho \vec{v} \vec{v}+\tens I\left(p+\frac{1}{2}|\vec{B}|^{2}\right)-\vec{B} \vec{B} \\
\vec{v} \vec{B}-\vec{B} \vec{v}+\underline{\tens I \phi} \\
\left(E+p+\frac{1}{2}|\vec{B}|^{2}\right) \vec{v}-\vec{B}(\vec{v} \cdot \vec{B}) \\
V_{r e f}^{2} \vec{B} \\
\end{array}\right)=\left(\begin{array}{c}
0 \\
\rho \boldsymbol{g}\\
0\\
\rho \boldsymbol{g} \cdot \vec{v} \\
0 \\
\end{array}\right).
\end{equation}
where, $\rho$ is the density, $P$ is the thermal gas pressure, $\vec g(r) = -(G M_\odot/r^2)\, \hat{\vec e}_r$ gravitational acceleration, and $\tens I = \hat{\vec e}_x \otimes \hat{\vec e}_x + \hat{\vec e}_y \otimes \hat{\vec e}_y + \hat{\vec e}_z \otimes \hat{\vec e}_z$ corresponds to the identity dyadic. The last equation in the set above corresponds to the conservation equation for magnetic flux for hyperbolic divergence cleaning. This way, we assure that the divergence of the magnetic field in the domain is negligible. The entire MHD formulation is normalised by the reference values of $\rho_\text{ref} = 1.67 \cdot 10^{-13}$ kg.m$^{-3}$, $B_\text{ref} = 2.2 \cdot 10^{-4}$ T, $l_\text{ref} = 6.9551 \cdot 10^{8}$ m and the corresponding $V_A$ computed from the values above. 

To create the surrounding field (i.e.\ without flux rope) the state variables are derived from a one-point backward differentiation scheme. In contrast to the steady-state solver setup introduced in \citet{Perri2022}, this study required a transformation into a time-accurate model in order to track the CME propagation. This was achieved by using a 3-point backward time discretization with a time-step limited to $1 \cdot 10^{-2}$ (in code units, i.e.\ $14.4\;$s in physical time). The resulting linearized system was solved using the Generalized Minimal RESidual (GMRES) method with a parallel Additive Schwartz preconditioner from the PETSc library \footnote{\url{https://petsc.org/release/}}. The time-accurate iterative process is considered converged at each time-step if the residual is lower than $1 \cdot 10^{-4}$ or after four iterations. We conducted numerous tests to determine the optimal values for these three parameters (time-step, maximum number of iterations and targeted residual) in order to achieve the best accuracy while minimizing numerical resources. In particular, we conducted tests using time steps that were ten times larger (i.e. $10^{-1}$). Unfortunately, this led to convergence issues, resulting in significant alterations in the simulation results during the propagation. We also attempted simulation using a time step that was ten times smaller (i.e. $10^{-3}$). There were no notable deviations in the results from what is observed in the simulations that employed a time step of $10^{-2}$ (cf. Sect. \ref{sec:Prop}). The only differences observed were small ($<5\%$) discrepancies in the magnitude of the profiles presented in section \ref{sec:profiles} between the $10^{-2}$ and $10^{3}$ time steps. However simulation that used a time step of $10^{-3}$ lasted more than three days. Hence, we concluded that the accuracy gains were insufficient to justify the allocation of significantly more numerical resources. It is worth noting that the limitations we encountered, such as the existence of high-speed streams (cf. Sect. \ref{sec:limitation}), persisted even when using a time step of $10^{-3}$.

With the final set of parameters, the simulations on the GENIUS cluster of the Vlaams Supercomputing Center\footnote{\url{http://www.vscentrum.be}} required between 20 and 26 hours of computation using 196 cores in parallel to reach 6000 iterations, i.e. the prescribed stop condition. After 6000 iterations, which equates to approximately 24 hours in physical time, in most of the simulations, the magnetic ejecta has successfully traversed the boundary at 0.1~AU (as discussed in Sect. \ref{sec:profiles}). The only exception is the slowest simulation with $\zeta=5$ in the minimum activity, where the magnetic ejecta is still in the process of crossing the boundary when the simulation ends. However, this simulation has still provided us with valuable information regarding the profiles of various magnetic and thermodynamic quantities, their maximum values, and the CME's geometry during its propagation (cf. Sects. \ref{sec:Prop} and \ref{sec:profiles}). It's also noteworthy that shorter simulation durations are sufficient for faster rope flux (e.g. 3750 iterations instead of 6000 in the simulations with $\zeta>3$ in the maximum activity solar wind). However, we chose to maintain uniformity in our results by employing the same simulation duration for all cases.

The boundary conditions are prescribed as following. The inner boundary is at $1\;R_{\odot}$. The density on the inner boundary is set to a constant value of $\rho_\text{ref} = 1.67 \cdot 10^{-13}$ kg.m$^{-3}$. According to the Parker's solar wind model, a small outflow from the surface is prescribed to follow the magnetic field lines \citep[see][for details]{Brchnelova2022b}, the radial component of which is determined based on the imposed magnetogram. The magnetograms used are obtained by the Helioseismic and Magnetic Imager (HMI) on board of the Solar Dynamic Observatory (SDO) satellite \citep{Scherrer11}, following the study of \citet{Perri23}. To couple the model with EUHFORIA, the outer boundary should be set at least at $21.5\;R_{\odot}$. However, according to the recommendations of \citet{Brchnelova2022b}, the grid extends beyond $21.5\;R_{\odot}$ to $25\;R_{\odot}$ in order to reduce the impacts of the outer boundary on the quantities that will be used in the heliospheric simulations (e.g.\ EUHFORIA). Figure 17 in \citet{Brchnelova2022b} provides a graphical depiction of the final grid used.  

\subsection{Solar wind configurations} \label{sec:validation}

\begin{figure*}[ht!]
\centering
\includegraphics[width=0.95\textwidth]{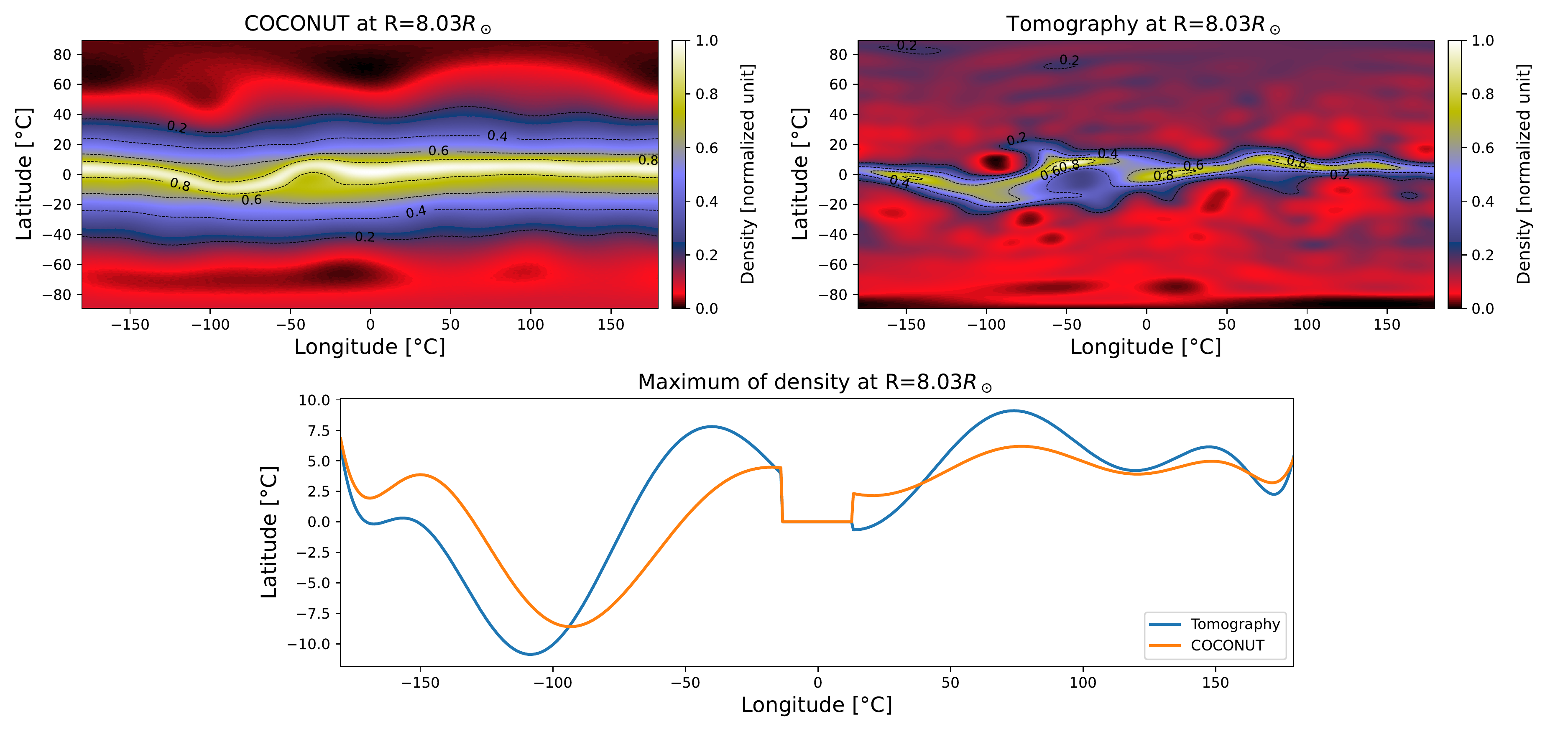}
\caption{Comparison of the density values derived from COCONUT and those obtained through the tomographic process. Top panels : density maps of the solar corona at height $R=8.0\;R_{\odot}$ for the total solar eclipse of $2^{nd}$ July 2019 in Carrington longitude–latitude coordinates. The left panel is the density extracted from the COCONUT model and the right panel is obtained from the tomography process detailed in the section \ref{sec:validation}. The color bar corresponds to the density in normalized unit. The dotted lines delineate isocontours for density values of [0.2, 0.4, 0.6, 0.8]. The bottom panel shows the position of the maximum of density in Carrington longitude–latitude coordinates. The orange curve is the maximum obtained from COCONUT while the blue one is for the tomography data. The position of the equatorial streamer belt which has been reconstructed by COCONUT is consistent with the one obtained by tomography. }
\label{fig:tomo}
\end{figure*}

The first step in our work is to model the solar wind in which the flux rope will propagate. This is achieved by running COCONUT in its relaxation mode as presented in \citet{Perri23}. Initially, as an initial condition, we compute a potential field approximation using a fast Finite Volume solver implemented within COCONUT for the Laplace equation \citep[cf. Sect. 2.3 in][]{Perri23}. Then, the global magnetic field distribution is obtained after the MHD relaxation.

Two magnetic configurations are considered for our further discussion. The first test-case is based on magnetograms from the solar eclipse of $2^{\rm nd}$ July 2019. The coronal model corresponds to a minimum of solar activity. The second test-case represents a maximum of solar activity from the solar eclipse of the $20^{\rm th}$ March 2015. Cases corresponding to total solar eclipses were originally chosen in order to optimize the comparison with observations \citep[cf.][]{Perri2022, Perri23, Kuźma23}.

The left panels of Fig.~\ref{fig:minmax} show the original pictures of the solar eclipse as obtained from 128 images in case of solar minimum (top-left, Druckm\"uller and Aniol 2019) and 29 images in case of solar maximum (bottom-left, Druckm\"uller et al 2015). The right panels illustrate numerical results which were obtained using COCONUT. The sphere color map represents the radial component of the magnetic field at the inner boundary, while white lines represent magnetic field lines of the magnetic coronal structures in the plane of sight.

Once the solar wind is modeled and before inserting the flux rope in the global coronal model, it is needed to validate the numerical scheme by comparing its output to real features and structures of the solar corona. It was revealed by \citet{Yeates18}, who compared seven different models applied to the same simulation of coronal magnetic field, that even for the same input dataset different models can produce considerably different results. In light of these findings, it is essential to compare obtained results with ground-based or in-situ observations every time a new model is introduced. It is especially important for simulations related to space weather forecast infrastructure, as even slight changes in inclination of magnetic structures may greatly affect the propagation of evolving CMEs at 1~AU and thus their geoeffectivity.

The validation process was already performed by \citet{Kuźma23} for our two solar wind configurations in COCONUT. The validation scheme, originally described in \citet{Wagner22}, is meant to compare fully relaxed, steady-state solutions with the global coronal structures observed in white light. It consists of a five-step process that includes visual classification, feature matching, streamer direction and width determination, brute force matching, and topology classification. 

In our study, the only difference with the simulations of \citet{Kuźma23} is that we employed a finer grid resolution. For our simulations, we employed a grid that ranged from R=1 $R_{\odot}$ to R=25 $R_{\odot}$, with a variable cell size between $10^{-8}$ and 0.215 $R^{3}$. In contrast, \citet{Kuźma23} used a grid that extended radially outward in layers between R=1 $R_{\odot}$ and R=21.5 $R_{\odot}$, with a cell size that varied from $10^{-7}$ to 0.365 $R^{3}$. Nevertheless, this does not affect the conclusions: in both solar wind configurations, the equatorial structures are correctly placed, shaped, and inclined (cf.\ Fig.~\ref{fig:minmax}). The apex of streamers matches, and the coronal holes are well-defined. The only area where a mismatch is observed is the north polar region in case of the maximum of the activity (cf.\ bottom panels in Fig.~\ref{fig:minmax}): while in the eclipse observations magnetic structures are present there, they are absent in numerical results. However, the aforementioned issue generally arises due to insufficient coverage of the polar regions by photospheric magnetic field observations. This, in turn, leads to inadequate resolution of coronal magnetic structures in that region. It should be noted that this issue does not usually arise during the minimum of solar activity. At this time, the magnetic configuration tends to be more dipolar in nature and with few small magnetic structures near the polar regions. Such a configuration is, therefore, easier to reconstruct even from low-quality observational data. Finally, we can conclude that our steady-state model correctly recreates global coronal structures and this is sufficient for studying CME propagation within its framework.

In addition to the validation performed by observing the global coronal structures in white light, we propose an additional comparison which is based on computing the electron density of the coronal plasma using a coronal rotational tomography process. Originally developed by \citet{Morgan15}, the tomography involves constructing a density map at different heights by using the polarized brightness observations provided by the STEREO A COR2 coronagraph over a half-Carrington rotation period. The tomography is based on advanced data processing and calibration detailed in \citet{Morgan15}, followed by a spherical harmonic-based regularized inversion method \citep{Morgan19} which was later improved by \citet{Morgan20}. Different tomography maps covering a broad range of years and heliocentric distance are publicly available online \footnote{\url{https://solarphysics.aber.ac.uk/Archives/tomography/}}.

Figure \ref{fig:tomo} compares the electron density map at $R=8\;R_{\odot}$ obtained by COCONUT (left) with that obtained by the tomography process (right) for the minimum activity simulation (i.e the solar eclipse of $2^{\rm nd}$ July 2019). The tomography map shows more detailed structure than COCONUT where the distribution of structure is along nearly constant latitude bands. The highest densities in tomography map are reached locally at longitudes equal to $-50^\circ$ , $15^\circ$ and $96^\circ$. In both cases the structure is close to an equatorial single narrow streamer sheet. However, the transition between the streamer belt and its surrounding regions is smoother in the COCONUT map than in the reconstructed density map. 

We have also observed that the equatorial structure's width in COCONUT is significantly greater than that of the tomography map. Specifically, the width measures approximately $\approx40^\circ$ in COCONUT, while it only reaches a maximum of $\approx20^\circ$ in the tomography map. This disparity in width is directly related to the resolution of the grid used in the simulations. \citet{Brchnelova2022b} found that the current sheet's shape and behavior are strongly influenced by the numerical dissipation resulting from the finite discretization of the system and the level of magnetic divergence present within the domain. Due to these two factors, premature reconnection occurs at an X-point, and the location of this point depends on the resolution. This early reconnection event causes an increase in the current sheet's width as it progresses from the X-point. It is noteworthy that even with an increase in the number of cells (from 1.5 million to 2.3 million), this non-physical behavior persists and is also observed in other solvers such as Wind-Predict \citep{Perri2022}.

To compare the positions of the streamer belt, the bottom panel of Fig.~\ref{fig:tomo} shows the latitude of the maximum density as a function of longitude for both COCONUT and the tomography process. The curves have been smoothed using a Savitzky-Golay filter \citep{Press90}. The discontinuity observed near longitude and longitude $0^\circ$ is a numerical artifact resulting from the method which was used to extract the maximum density. In Fig.~\ref{fig:tomo}, it can be seen that the dynamics of the two curves are almost identical. However, the COCONUT result appears slightly delayed compared to the tomography density before the longitude $0^\circ$. Additionally, the variations in the tomography density are wider (cf.\ Fig \ref{fig:tomo}, right panel). On average, the difference between the two curves is only $2.1^\circ$, with a maximum deviation of $6.7^\circ$. Despite the simplifying assumptions of the COCONUT model, which is only polytropic (cf.\ Sect.~\ref{sec:Coconut}), the density is in good agreement with the observations. This agreement could be improved by incorporating more source terms in the COCONUT MHD equations, but this is beyond the scope of this work.

The comparison between the density of COCONUT and the tomography process has also been performed at multiple heights (4, 4.4, 5, 5.4, 5.9, 6.5, 6.9, 7.5 and 8 solar radii) to have more confidence in the validity of the MHD results. However, the structure is almost identical between heights as expected in the inner corona.

For the maximum activity studied in our work no tomography map is available. The rapid changes related to the large CME occurring during this high period of activity lead to significant error in the 'static' tomography method. Nevertheless, the results obtained for the minimum activity, along with the comparison with the white light images, provide sufficient validation for the coronal magnetic field.

\section{Flux rope in COCONUT} \label{sec:Fluxincoconut}
\subsection{Titov-Démoulin modified model} \label{sec:TDM}

\begin{table}[ht!]
\centering
\begin{tabular}{cccc}
 & \multicolumn{3}{c}{Initial parameters} \\ \hline
$\zeta$ &  {$B_{0}$ {[}G{]}} &  {$F$ $[10^{21} Mx]$} & $V_{0}$ $[km/s]$ \\ \hline
5 &  {4.2} &  {1.6} &  {348}   \\ 
6 &  {5.1} &  {2.1} &  {431}   \\ 
7 &  {6.0} &  {2.5} &  {493}    \\ 
8 &  {6.9} &  {2.9} &  {572}   \\ 
9 &  {7.8} &  {3.3} &   {638}   \\ 
10 &  {8.7} &  {3.6} &  {707}   \\ 
11 &  {9.6} &  {4.0} &  {752}   \\ 
12 &  {10.5} &  {4.4} &  {827}   \\ 
13 &  {11.4} &  {4.8} &  {877}    \\ 
14 &  {12.3} &  {5.2} &  {928}   \\ 
15 &  {13.2} &  {5.5} &  {990}   \\ 
16 &  {14.1} &  {5.9} &  {1035}    \\
17 &  {15.0} &  {6.3} &  {1067}   \\
18 &  {15.9} &  {6.7} &  {1130}   \\
19 &  {16.8} &  {7.0} &  {1189}   \\
20 &  {17.7} &  {7.4} &  {1220}   \\ 
\end{tabular}
\caption{Summary table of the different simulations studied at minimum of activity. The main column named "Initial parameters" shows the initial magnetic field of the flux rope ($B_{0}$), the initial magnetic flux ($F$), and the initial velocity ($V_{0}$) in function of the parameter $\zeta$. $V_{0}$ is extracted at the time of the first saved time step, i.e. $t=0.08h$.}
\label{tab:minimum}
\end{table}

\begin{table}[ht!]
\centering
\begin{tabular}{cccc}
 \multicolumn{4}{c}{Initial parameters} \\ \hline 
 $\zeta$ & {$B_{0}$ {[}G{]}} &  {$F$ $[10^{21} Mx]$} & $V_{0}$ $[km/s]$ \\ \hline
 2 &  {9.8} &  {4.1} &  {455}\\ 
 3 &  {14.5} &  {6.1} &  {772} \\ 
 4 &  {19.4} &  {8.1} &  {1060} \\ 
5 &  {24.4} &  {10.2} &  {1341} \\ 
6 &  {29.5} &  {12.3} &  {1617} \\ 
7 &  {34.6} &  {14.5} &  {1888} \\ 
8 &  {39.7} &  {16.7} &  {2146} \\ 
9 &  {44.8} &  {18.8} &  {2385} \\ 
\end{tabular}
\caption{Same as Fig.~\ref{tab:minimum} but for the simulations with the maximum of activity configuration. }
\label{tab:maximum}
\end{table}

The CME model implemented in COCONUT is an analytical circular flux rope originally developed by \citet[][hereafter TD]{Titov99} and then modified by \citet[][hereafter TDm]{Titov14}. The model depicts a circular cross-section with a current-carrying, approximately force-free magnetic field in the toroidal segment. It is partly immersed in the photosphere, such that only one circular arc protrudes into the corona.

In the original TD model, the current density is nearly uniform along the axis of the flux rope. Instead, in its modified version, the current density can be either (1)~mainly localized close to the boundary or (2)~parabolically distributed within the cross-section with a maximum at the axis and cancelling at the boundary. As a result, the twist can either be concentrated in a thin layer or distributed over the whole torus.

The construction of the flux rope in the model is based on the distribution of the ambient field, which must be potential in the surrounding area. Specifically, the flux rope is placed such that its plane is locally perpendicular to the surrounding fields along one of the approximately circular isocontours. The magnetic field of this isocontour perpendicular to the toroidal axis is denoted as $B_{\perp}$. In this configuration, the flux is in equilibrium when the magnetic pressure resulting from the net current $I\neq 0$ is balanced with the tension generated by the ambient field. In other words, the intensity must be equal to the Shafranov intensity such as :
\begin{equation} \label{eq:IS}
  I\equiv I_{S}\approx \frac{4\pi RB_{\perp}/\mu}{\ln{\frac{8R}{a}}-\frac{3}{2}+\frac{I_{i}}{2}},
\end{equation}
where $R$ is the radius of the torus, $a$ the minor radius and $I_{i}$ the internal self-inductance per unit length of the rope. The curvature of the flux rope is characterised by the ratio between the small radius and radius of the torus : 
\begin{equation}
  \epsilon=\frac{a}{R}\ll 1.
\end{equation}
In this configuration, the axial magnetic flux is defined as : 
\begin{equation} \label{eq:F}
  F\equiv F_{S}\approx \mp\frac{1}{2}\mu I a, 
\end{equation}
for the density distribution number (1), and 
\begin{equation}
  F\equiv F_{S}\approx \mp\frac{3}{5\sqrt{2}}\mu I a. 
\end{equation}
for the case (2). In the above equations, the sign is positive if the axial field and the current are counter-directed and negative otherwise.

Assuming an axial symmetry of the magnetic field in the torus, it is possible to derive the magnetic vector potential from the current and the magnetic field defined above \citep[cf.\ Sect.~2 in][]{Titov14}. Finally, we note that all the characteristics of the flux rope can be determined using only the radius $R$, the minor radius $a$ and an isocontour of the ambient magnetic field. 

\subsection{Implementation in the solar wind} \label{sec:implementation}

\begin{figure*}[t]
\centering
\includegraphics[width=0.80\textwidth]{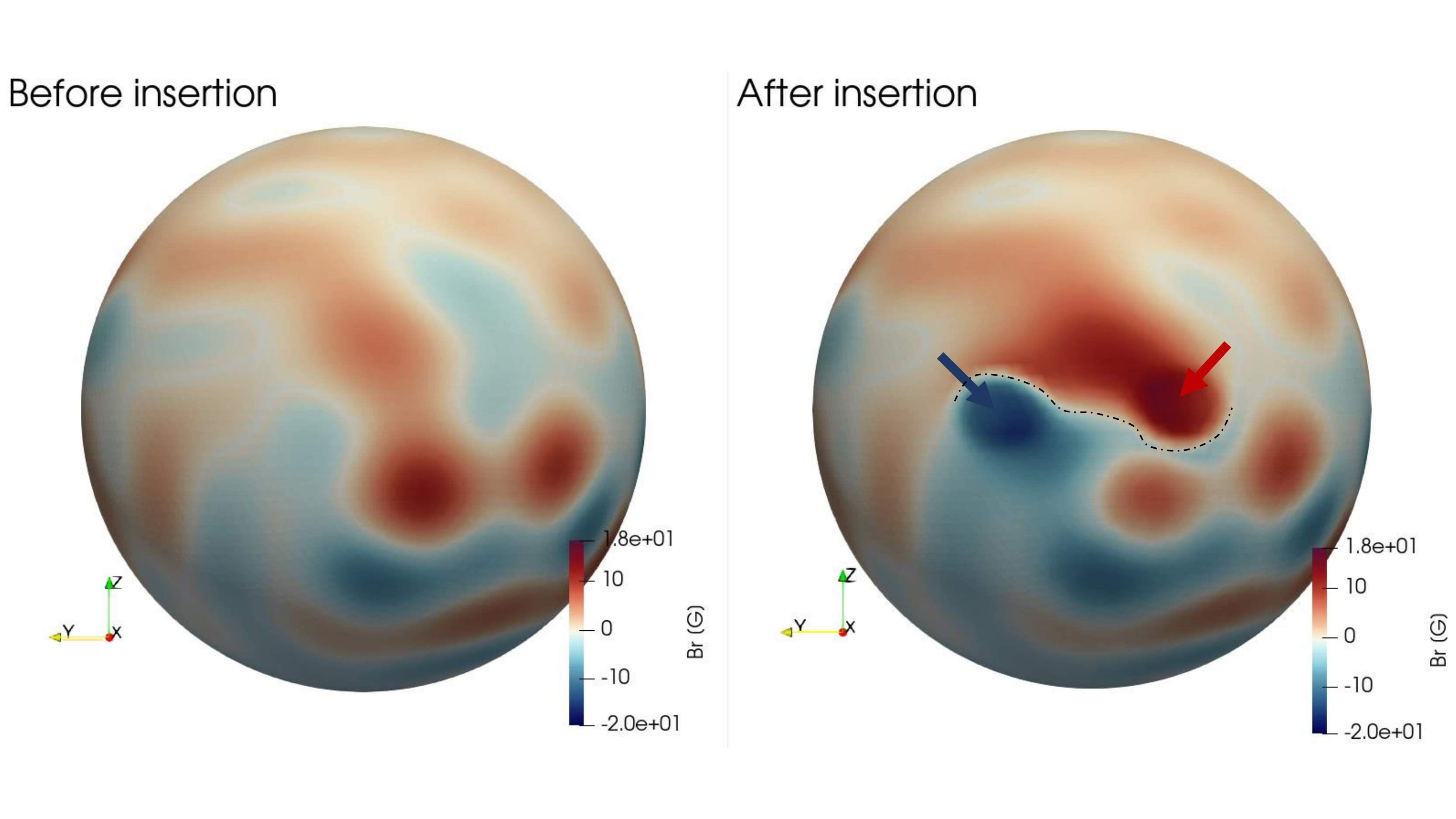}
\caption{Snapshots of the radial magnetic field $B_r$ at $R=1\;R_{\odot}$ in the COCONUT model (lower radial boundary corresponding to the base of the solar corona). The left panel shows the magnetic field produced by the initial relaxation run for the maximum activity solar wind, while the right panel shows the same location after the insertion of the flux rope. The insertion of the flux rope results in the addition of two polarities and a modification of the near photospheric field. The blue arrow indicates the negative polarity, while the red one indicates the positive polarity. The magnetic inversion line is demarcated by the dotted line.}
\label{fig:photosphere}
\end{figure*}

As previously discussed (cf.\ Sect.~\ref{sec:TDM}), the flux rope is in equilibrium when its intensity is equal to the Shafranov intensity (cf.\ Eq. \ref{eq:IS}). However, in this work, the intensity of the ring current is deliberately set higher than the Shafranov intensity to produce an eruptive behavior. Therefore, the intensity of the ring current is defined as:
\begin{equation}
  I=\zeta I_{s}
\end{equation}
with $\zeta$ a positive number. By setting $\zeta>1$, the flux rope is unstable from the start of the simulation. The overlying background magnetic field can not prevent the outward expansion of the flux rope. As a result, there is no relaxation phase in which the photospheric patterns (e.g.\ shearing, twisting or emergence of polarities) drive the magnetic system until the trigger of eruption due to magnetic reconnection or an ideal instability \citep{Green18}. 

Our simulation process is carried out in two steps. First, the solar wind background is created by a relaxation run (cf.\ Sect.~\ref{sec:validation}). Once this is done, the flux rope is inserted into it and COCONUT is restarted in its time-dependent version. During the insertion, only the magnetic field in the domain is modified. The density, the velocity and the pressure are not changed. To insert the flux rope numerically, we use the python library called pyTDm developed by \citet{Regnault23} and available online\footnote{\url{https://github.com/fregnault/pyTDm}}. Originally developed for the PLUTO solver, the module has been adapted to work with COCONUT data files. The modification is also publicly available online.

The code pyTDm is designed to work only with structured mesh. As a result, an interpolation step is required to convert the unstructured COCONUT grid to the right format. The radial basis function (RBF) interpolation is the method used for this step \citep{Buhmann00}. This method produces accurate results even for grids with many cells. The choice of the interpolation method and its parameters was optimized to ensure that the solar wind produced by COCONUT after the relaxation run is only slightly modified during this procedure.

The module pyTDm offers the ability to specify the initial position of the flux rope (latitude, $lat$, and longitude, $lon$ ), the radius $R$, the minor radius $a$, the depth at which the torus is buried below the photosphere $d$ (i.e.\ the distance between the surface of the Sun and the center of the torus), the tilt and the parameter $\zeta$. The current density is distributed along the case (1) described in section \ref{sec:TDM}, i.e.\ mainly in a thin layer close to the boundary.

In \citet{Regnault23}, they defined specific boundaries conditions around the flux rope at $R=R_{\odot}$ that were different from those of the solar wind boundary conditions. In contrast, in this study, there is no such distinction. It is also important to note that the insertion of the flux rope structure will significantly change the photospheric magnetic field. This results in the addition of two opposing magnetic polarities as can be seen in Fig.~\ref{fig:photosphere}, which illustrates the modification of the photospheric magnetic field after the injection of a flux rope in the minimum activity configuration. The TDm is not positioned at the location of an observed active region, meaning that the CME inserted does not correspond to a CME that was possibly observed on the day of the photospheric magnetic field measurement. 

In Fig.~\ref{fig:photosphere}, we can also see an inversion line between the two polarities. This reflects the local presence of potential magnetic fields that overlay the flux rope. The magnetic field, a bit further away from the insertion zone also seems to be modified. This is particularly noticeable in the left sunspot below the positive polarity. The magnetic field of this sunspot is lower after the injection of the flux rope.

Finally, we can mention that that the disparities observed in Fig.~\ref{fig:photosphere} are the only contrasts between the photospheric magnetic field after and before its insertion.
The areas that are not covered by the figure (mainly the opposite side) are identical. This means that the interpolation step, that takes place after the execution of the pyTDm module in order to put the magnetic field configuration back into the unstructured grid of COCONUT, does not impact the photospheric field and that the modification is therefore mainly related to the insertion of the flux rope. 

\subsection{Initial parameters} \label{sec:initial}

\begin{figure}[h!]
\centering
\includegraphics[width=0.5\textwidth]{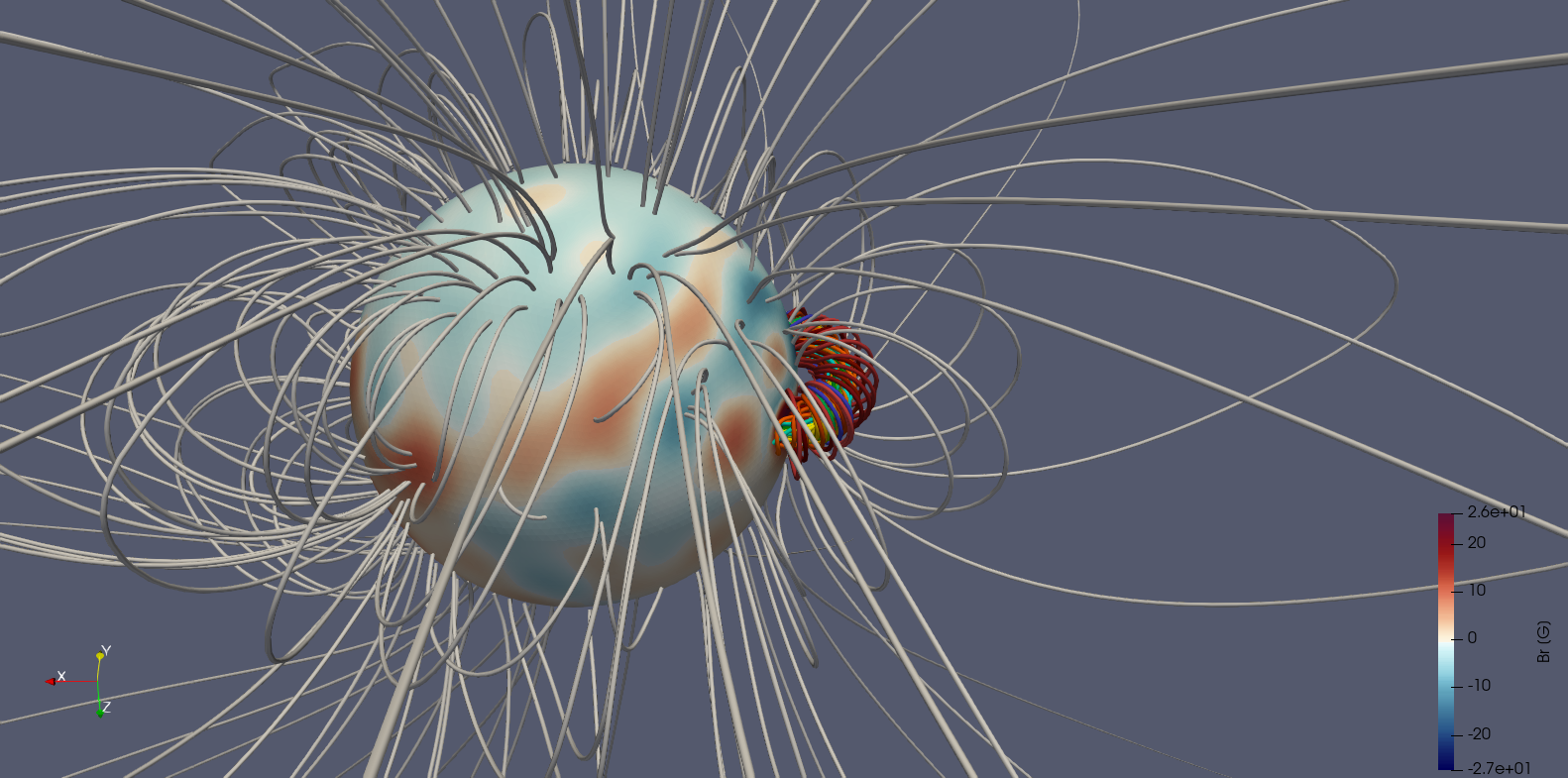}
\caption{Vizualisation of the Titov-Démoulin modified flux rope in COCONUT in the maximum activity solar wind. The colored lines correspond to different magnetic field lines. The seed to trace the field lines is a sphere of radius $0.1\;R_{\odot}$ which is placed at the positive polarity. The grey field lines are a sample of the surrounding field. The radial magnetic field at the solar surface (in Gauss) is also displayed. The flux rope is in the equatorial plane and initially only an arc extends from the surface of the Sun.}
\label{fig:torus}
\end{figure}

In this study, simulations were conducted to investigate the effect of initial conditions on the properties of flux ropes at 21.5 solar radii. A total of 24 simulations were run: 16 in a low-activity phase surrounding magnetic field and 8 in a high-activity phase field.

All the flux ropes are placed in the equatorial plane at longitude $lon=0^\circ $ and latitude $lat=90^\circ$. Their centres are offset by $d=0.15\;R_{\odot}$ from the solar surface. Their major radius are $R=0.3\;R_{\odot}$ and their minor radius $a=0.1\;R_{\odot}$. This results in a polarity area of $A=4839~Mm^{2}$ on the inner boundary. The chosen size is deliberately larger than the typical size of an active region to ensure adequate numerical resolution for the description of the CME and to avoid interpreting it as noise. As shown in Fig.~\ref{fig:torus}, after its insertion in the maximum activity solution generated by COCONUT, with the selected parameter set, only one arc protrudes into the solar corona.

For a given phase of the solar cycle (active or quiet), the difference between all the simulations is only due to variations in the parameter $\zeta$, which affects the intensity of the initial flux rope. Tables \ref{tab:minimum} and \ref{tab:maximum} summarize the different cases examined for the minimum and maximum of activity. In the first magnetic configuration, the range of $\zeta$ varies from 5 to 20. The magnetic field of the flux rope at its axis being directly related to the intensity, it increases from $4.2\;$G to $17.7\;$G. The magnetic field and the parameter $\zeta$ are related by the equation $B_{0}(\zeta)=0.9 \zeta - 0.3$. This linear relationship is a result of using the same surrounding magnetic field as a reference for all runs (cf. Sect.~\ref{sec:implementation}).

For the maximum activity configuration, the ambient magnetic field is stronger at the same height. Therefore, when the parameter $\zeta$ ranges from $2$ to $9$, the magnetic field increases from $9.8\;$G to $44.8\;$G. The relationship between the two quantities is: $B_{0}(\zeta)=5.1 \zeta - 1.1$. As will be explained in section \ref{sec:firststage}, the flux ropes with high magnetic flux lead to large initial speed. To avoid convergence problems with large velocity gradients, in the maximum activity solar wind, the parameter $\zeta$ does not exceed 9. In contrast, in the minimum activity cases, the $\zeta$ parameter does not go below 5 to maintain sufficiently high initial velocity. 

From the magnetic field, it is possible to compute the magnetic flux according to the equation~\ref{eq:F}. For the minimum activity, the magnetic flux varies from $1.6$ $10^{21}$ Mx to $7.4$ $10^{21}$ Mx. While for the maximum configuration, the flux is between $4.1$ $10^{21}$ Mx to $18.8$ $10^{21}$ Mx. The high values computed for the maximum activity correspond to major and rare solar events, the magnetic flux in the case of a minimum activity is closer to the values measured for large sunspots \citep{Van15}.

\section{Propagation in the corona} \label{sec:Prop}
\subsection{Early stages of the evolution} \label{sec:firststage}

\begin{figure*}[h!]
\centering
\includegraphics[width=1\textwidth]{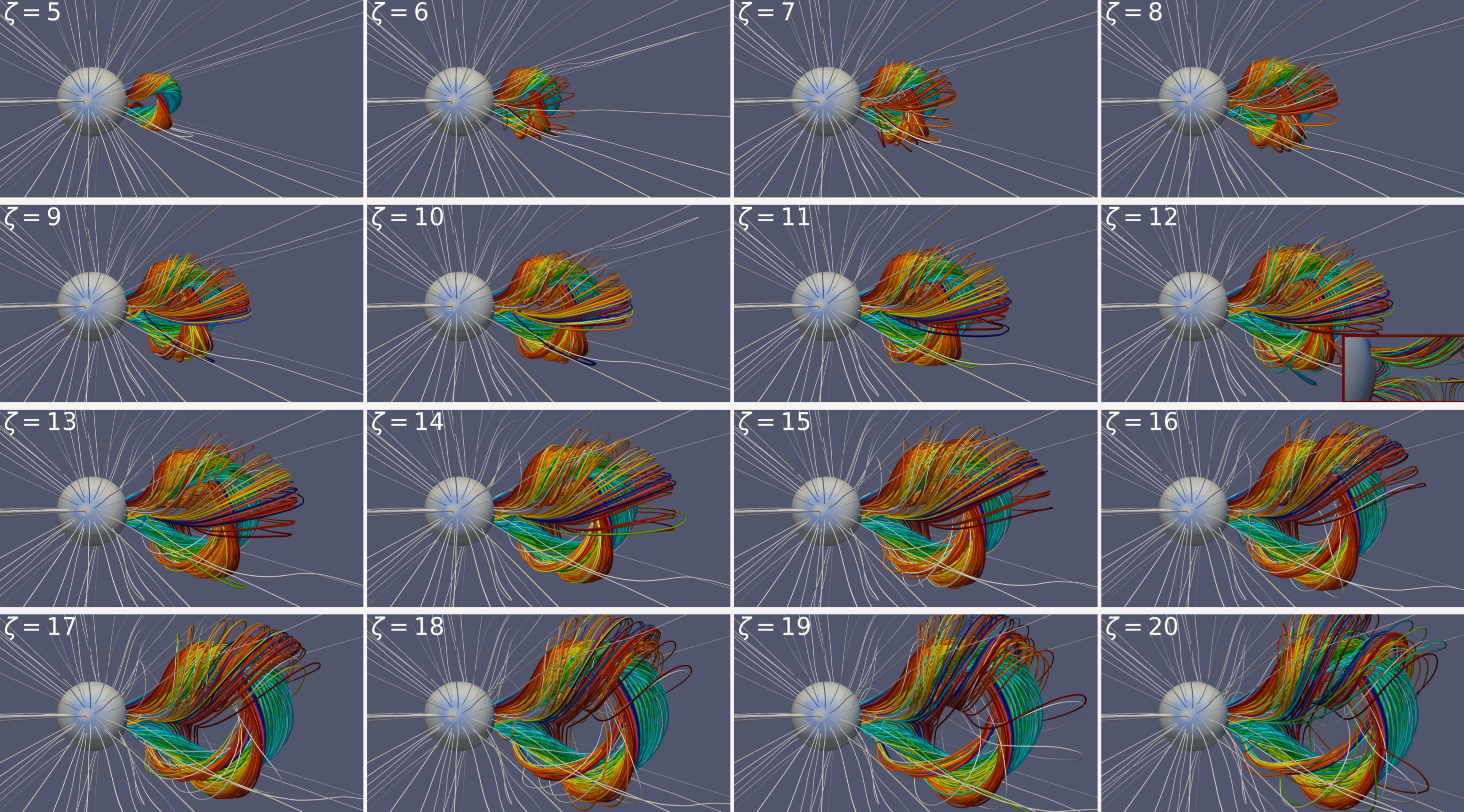}
\caption{Visualisation of all the flux ropes modelled in the solar wind reconstructed from a minimum of activity at the physical time $t=1.2\;$h. The magnetic field lines are displayed along the legend explained in the Fig.~\ref{fig:torus}. The small panel in the bottom right corner of the simulation "$\zeta=12$" is a zoom at the polarities highlighting the presence of post-flare loops.}
\label{fig:evol_min}
\end{figure*}

\begin{figure*}[h!]
\centering
\includegraphics[width=1\textwidth]{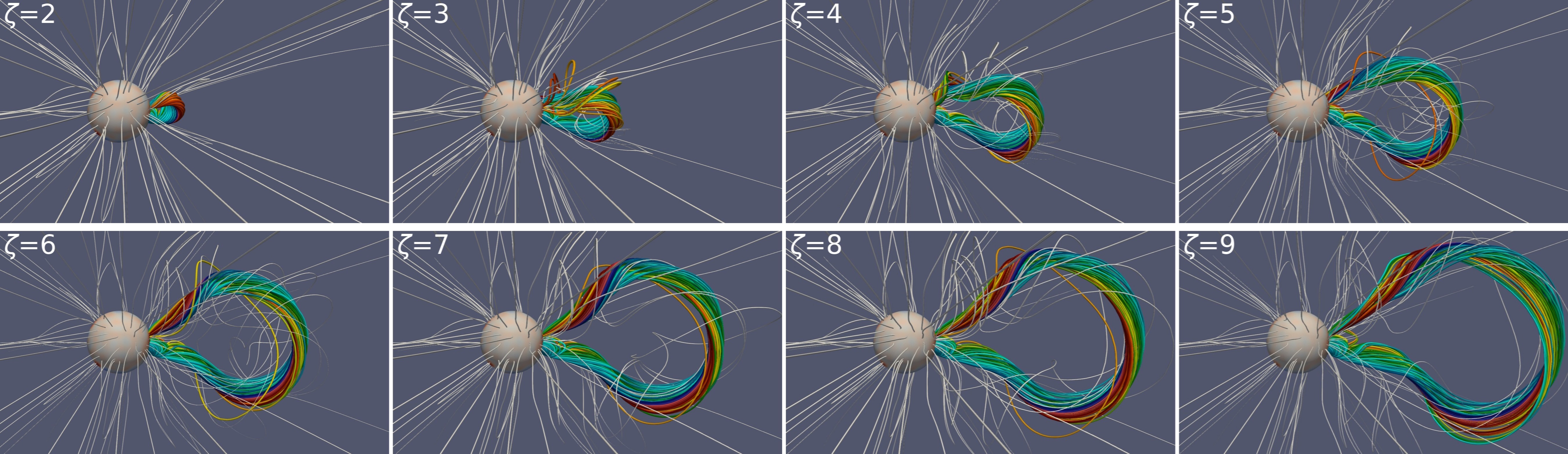}
\caption{Same as the Fig.~\ref{fig:evol_min} but for the simulations in the maximum activity solar wind configuration.}
\label{fig:evol_max}
\end{figure*}

\begin{figure*}[t]
\centering
\includegraphics[width=1\textwidth]{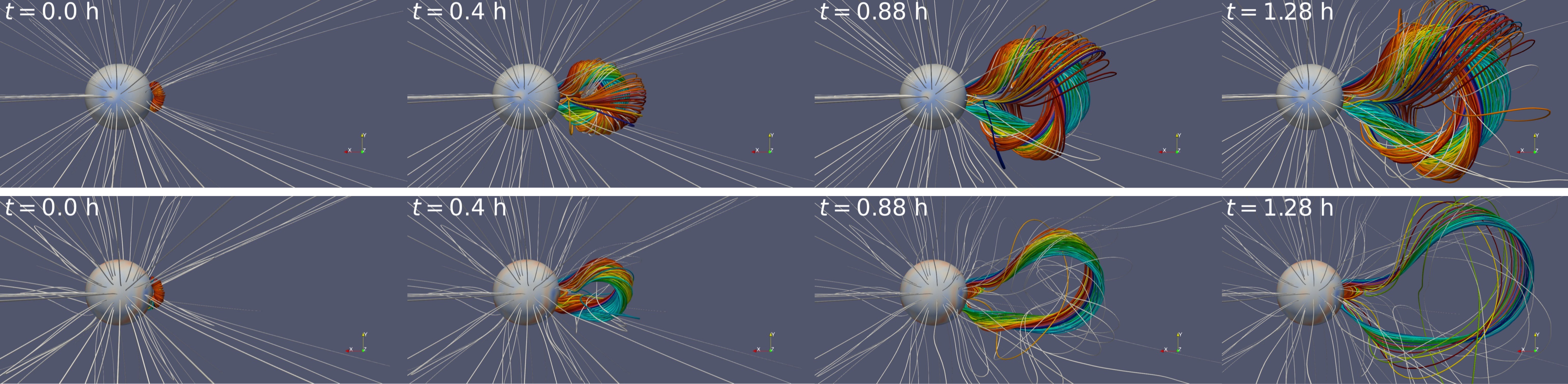}
\caption{Visualisation of the propagation of the TDm model in the solar wind reconstructed from a minimum of activity (top row) and from a maximum of activity (bottom row). For the two magnetic configurations, snapshots are taken at the beginning of the simulation and at physical times equal to $0.4\;$h, $0.88\;$h and $1.28\;$h. The legend is the same as the Fig.~\ref{fig:torus}. The geometry of flux ropes evolving in the maximum activity is not directly identical to the one observed in the minimum activity meaning the solar wind configuration has an impact on the geometrical properties.}
\label{fig:comp_evol}
\end{figure*}

In order to track the propagation of all our flux ropes, we saved the 3D magnetic field and plasma flow outputs of COCONUT every 20 iterations corresponding to a time step of $dt=0.08\;$h, where $t$ is the physical time in the simulation. All the cases were stopped when the physical time exceeded $24.04\;$h.

 Figs. \ref{fig:evol_min} and \ref{fig:evol_max} show the flux ropes at the time $t=1.2\;$h in the solar wind obtained from the minimum and the maximum activity, respectively. The different magnetic field lines of the torus are depicted with different colors and originate from a sphere of radius $0.1\;R_{\odot}$ which is situated at the feet of the flux rope with a positive polarity. It is worth mentioning that we have tested different choices of seed for the magnetic field lines but this did not change the main results presented in this section. We also tried adding a sphere at the negative polarity. However, the increased number of magnetic field lines made the figure less clear and did not provide any additional information. Finally, the sphere at only the positive polarity was deemed the most appropriate for portraying the flux rope geometry. Additionally, a sample of the surrounding magnetic field is also shown in white. The only variation between the panels in each figure is the chosen parameter $\zeta$ during the implementation of the flux rope (cf.\ Sect.~\ref{sec:implementation}).

First, we see that the flux ropes in all simulations exhibit a self-similar evolution primarily in the equatorial plane. However, they do not reach the same height at the same instant. For a given solar wind, the higher $\zeta$ value results in a larger travel distance for the front of the flux rope. As expected, flux ropes with the highest $\zeta$ (i.e, highest intensity, cf.\ Eq. \ref{eq:IS}) have the highest initial speed. Tables \ref{tab:minimum} and \ref{tab:maximum} show the radial velocity of the top part of the CMEs in the minimum and maximum activity respectively, at the first time step saved, $t=0.08\;$h. It should be noted that while comparing velocity at $t=0.08\;$h, may not be entirely accurate, as the flux ropes have not reached the same height, it is still a good approximation. It is important to note that the initial velocity are not numerically prescribed, the latter is exclusively generated by the imbalance of Lorentz forces that occurs right from the beginning of the simulations \citep[cf.\ Sect 2.3 in][for more details on role of the Lorentz force on CME propagation]{Scolini19}.

At minimum of activity (cf.\ Tab. \ref{tab:minimum}), the initial radial speed ranges from $V_{0}=348$ km/s for $\zeta=5$ to $V_{0}=1220$ km/s for $\zeta=9$. While the simulations at maximum of activity cover a wider range of CME speeds, from $V_{0}=455$ km/s for $\zeta=2$ to $V_{0}=2385$ km/s for $\zeta=9$ (cf.\ Tab. \ref{tab:maximum}). The radial propagation speed of observed CMEs typically varies between $\sim200~km/s$ and $\sim2000~km/s$ \citep{Chen11}. The majority of our simulations fall within this range. Only the two most unstable cases in the maximum activity have higher speeds, but they can be considered as rare and fast events that are only occasionally observed. 

The higher the magnetic flux of the flux rope, the higher its initial CME speeds for a given solar wind. However, the velocity is not solely determined by the magnetic flux as evidenced by the fact that the simulation with $\zeta=11$ in the minimum activity and the simulation with $\zeta=3$ in the maximum activity do not have the same propagation speed despite having a similar magnetic flux ($\approx4.0-4.1$ $10^{21}Mx$). The density distribution and the surrounding field also play a role. There are various formulations to determine the propagation speed based on the characteristics of the solar corona. For instance, \citet{Shibata01} derived an analytical solution that depends on the Alfvén speed of the corona, and the density of both the corona and the flux rope.

Regarding the topology of the flux ropes, there are no significant differences in geometry when comparing simulation with a given $\zeta$ to the simulation with the subsequent $\zeta$ value (e.g.\ $\zeta=5$ and $\zeta=6$, $\zeta=7$ and $\zeta=8$ etc.) for the same solar wind. The increase in magnetic flux does not have a major impact on the geometry of the flux rope.

 Figures \ref{fig:evol_min} and \ref{fig:evol_max} show common points in the topologies of flux ropes as they grow into the solar corona. As they rise, the potential field lines surrounding it converge and pinch below the core of the flux rope. The convergence of these upward and downward field lines leads to the creation of a current sheet. This region is particularly suitable for magnetic reconnection. However, since there is no resistivity in the set of equations solved by COCONUT, the rearrangement of the field lines is done via numerical dissipation in our cases (cf.\ panel $\zeta=12$ in Fig.~\ref{fig:evol_min}). 

Reconnection leads to the formation of post-flare loops below the flux rope as seen in observations \citep{Schmieder95}. These field lines are located close to the polarities and are rapidly closing on the surface of the Sun. In addition to the post-flare loops, some field lines wrapping around the flux rope are created, although they may be not clearly visible in Figs. \ref{fig:evol_min} and \ref{fig:evol_max} because of the choices made for the visualization. Finally, we note that the rearrangement of the line-tied magnetic field lines should remove some constraint on the overlying field and thus facilitate the expansion of the flux rope.

Not all magnetic field lines are entirely rearranged. One part of the overlying magnetic loop is stretched upward by the advancing flux rope. The dynamics described above (e.g.\ the presence of post-flare loops), are described in the 2D "standard model" for CME originally developed by \citet{Carmichael64,Sturrock66,Hirayama74,Kopp76} and later extended in 3D by \citet{Aulanier12,Aulanier13,Janvier13}. Their model also accounts for physical processes not included in the COCONUT model, such as the presence of flare ribbons, the conversion of magnetic energy, and particle transport. COCONUT primarily focuses on the global distribution of magnetic field density in the solar atmosphere, rather than the processes leading to flare appearance.

Comparing Figs. \ref{fig:evol_min} and \ref{fig:evol_max}, we see that the geometry of the flux rope is not the same in the two solar wind configurations. To further emphasize these differences, the Fig.~\ref{fig:comp_evol} shows two magnetic flux ropes that propagate at approximately the same initial speed in the minimum and maximum activity configurations. For the minimum of activity, the CME has a $\zeta$ parameter of $17$ and an initial speed of $1067$ km/s, while in the maximum of activity the $\zeta$ parameter is $4$ and the initial speed is $1060$ km/s. Snapshots are taken at the first four time steps: $t=0\;$h, $t=0.4\;$h, $t=0.88\;$h and $t=1.28\;$h.

Initially, there is no noticeable difference between the two twisted flux ropes. As noted in section \ref{sec:implementation}, they have the same initial size and they are inserted at the same location on the Sun's surface, with only the background solar wind differing between them. At time $t=0.4\;$h, the flux ropes have reached almost the same height, which is expected given their nearly identical initial speed. However, their shapes differ.

At the times $t=0.88\;$h and $t=1.28\;$h, the geometry of flux ropes is significantly different. In the case of the flux rope evolving in the minimum of activity solar wind (cf.\ Fig.~\ref{fig:comp_evol}, top panels), a clear symmetry is observed along the $Z-X$ plane passing through the center of the torus. The difference between the two flanks can be numerically attributed to the fact that the origin of the magnetic field lines is a sphere placed at only the positive polarity. The geometry of the torus in this case is very close to the one observed by \citet{Regnault23} for flux ropes evolving in a dipolar ambient magnetic field. This similarity is expected as the ambient field during a minimum of activity is typically close to a dipole field (cf.\ Fig.~\ref{fig:minmax}). The results produced by the COCONUT and PLUTO solvers are thus consistent, despite the differences in their numerical scheme and grid resolution.

On the other hand, in the maximum of activity (cf.\ Fig.~\ref{fig:comp_evol}, bottom panels), the symmetry along the meridional plane crossing the front part of the flux rope is less pronounced. The ambient magnetic field surrounding the flux rope is not the same on either side of it. For instance, the positive polarity is in particular close to two important sunspots (cf.\ Fig.~\ref{fig:photosphere}). As the background magnetic field varies around the two legs of the flux rope, the interactions via magnetic reconnection between the field lines are also different. Additionally, the magnetic tension exerted by the surrounding field on the flux rope is not uniform along its length. The combination of these two factors leads to asymmetric evolution of the two legs.

We also observe that the magnetic flux rope appears thinner in the maximum activity than in the minimum activity. This difference comes from our choice of seed for tracing the field lines. In both cases and at all times, we represent only magnetic field lines crossing a sphere located at the positive polarity. However, for the simulations in the maximum activity, some of the magnetic field of the CME no longer pass through the original sphere. This is something that \citet{Regnault23} have also observed in their simulations. Indeed, after a few hours, the feet of the flux rope have slightly moved on the photosphere. The rearrangement of the field lines that compose them may lead to a shift in the anchoring area of the flux rope legs.

Further analysis, outside of the scope of this study, is needed in order to determine more qualitatively and quantitatively changes of geometry due to a different configuration of the solar wind; as well as the impact of visualization choices on the observed geometry. However, our first results suggest that the geometry of the flux is impacted by the surrounding magnetic field. In the section \ref{sec:profiles}, we will investigate the extent to which this difference in geometry affects the dynamics and values of the various magnetic and thermodynamic quantities at 21.5 solar radii.

\begin{figure*}[ht!]
\centering
\includegraphics[width=1\textwidth]{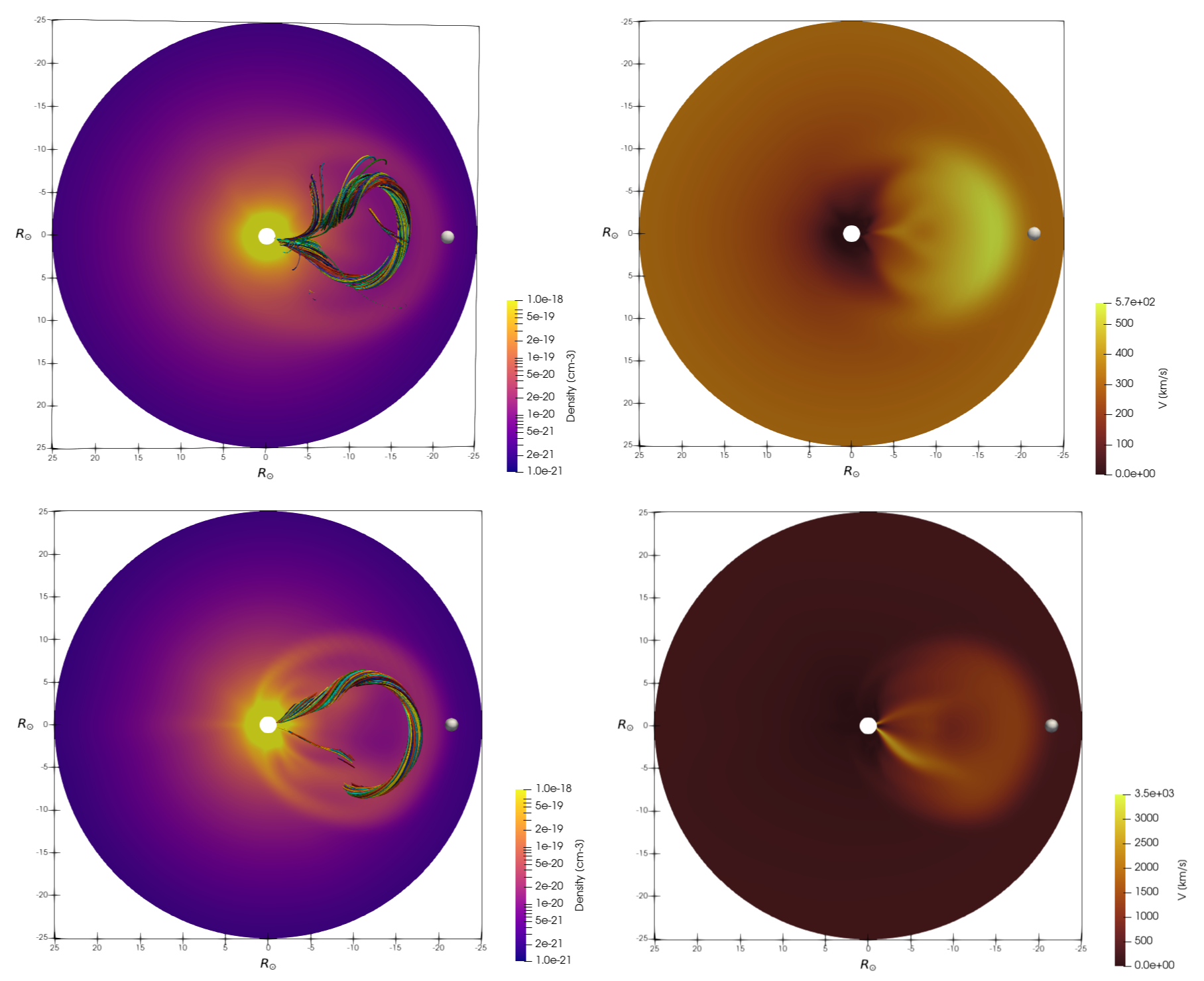}
\caption{2D equatorial slices of the density in log scale (left column) and of the radial velocity (right column). The top panels show the flux rope with $\zeta=15$ evolving in the minimum activity configuration while the bottom panels correspond to the simulation with $\zeta=9$ in the maximum activity. In the density panels, some magnetic field lines crossing a sphere of radius $1.5\;R_{\odot}$ located at $x=-16\;R_{\odot}$ are also displayed. Each color corresponds to one field line. A virtual satellite is placed at $21.5\;R_{\odot}$ ($X=-21.5\;R_{\odot}$, $Y=0$, $Z=0$), shown as a white sphere. The shape of the sheath is delimited by an ellipse in the density panels. In both solar wind configurations a sheath is ahead of the flux rope which is consistent with the observations. When the initial speed is too high, there are unphysical high speed streams in the wake of flux rope legs. }
\label{fig:density}
\end{figure*}

\subsection{Known limitations from high initial speed} \label{sec:limitation}

After the early stages of the evolution described in the previous section (cf.\ Sect.~\ref{sec:firststage}), the flux ropes continue to propagate in the solar corona. This propagation will impact all the MHD plasma and magnetic variables in the simulation domain. Fig.~\ref{fig:density} shows the density and the radial velocity in the equatorial plane for two very distinct simulations when the flux rope reaches around 15 $R_{\odot}$. The top panels correspond to the time $t=3.44$ h for the flux rope propelling with an initial velocity of approximately $990$ km/s in the minimum activity, while the bottom panels show the results obtained at the time $t=1.12$ h from the flux rope with the initial velocity of around $\sim2385$ km/s in the maximum activity. In the 2D slice of the density (cf.\ Fig.~\ref{fig:density}, left panels), some field lines of the flux rope crossing a sphere of radius $R=1.5\;R_{\odot}$ at $x=-16\;R_{\odot}$ are also displayed. To improve clarity, we have chosen not to display all 24 simulations. The features detailed above concerning density and the magnetic structures are valid for all cases studied for a given state of solar activity. Concerning the radial velocity, the right panels in the Fig.~\ref{fig:density} show examples for an extreme case (bottom panel) and for the more common case (top panel). 

First, the density distribution is affected in the same way in both solar wind configurations. In the opposite direction to the flux rope's propagation, the density decreases with distance from the Sun. In the direction of propagation, there is an increase in density by a factor of 10 along an almost circular band surrounding the flux rope. As the flux rope travels, it does not allow enough time for the solar wind to flow around it, resulting in an area of heated and compressed solar matter called the sheath. This area is turbulent and characterised by a higher density than the surrounding solar wind. It is worth noting that the accumulation of matter ahead of the flux rope is a process that enhances as the CME expands \citep{Siscoe08}. The amount of density reached in the sheath will be discussed later in the section \ref{sec:max_ftsolarwind}. The accumulation of density can be observed by in-situ spacecraft if the CME is faster than the solar wind \citep{Regnault20}. This confirms the ability of COCONUT to reproduce features observed during the propagation of a magnetic ejecta.

Regarding the magnetic structure (cf.\ Fig.~\ref{fig:density} left panels), no matter the solar wind and the initial velocity, it propagates radially in the equatorial plane. This suggests that there is no rotation of the flux rope. This is also the case for all simulations regardless of the distance from the sun. \citet{Regnault23} studied the propagation of the TDm model with the same initial geometry as our cases in a dipolar solar wind and they found that the flux rope rotates by almost $55^\circ$ before 15 solar radii. On the contrary, for thinner flux ropes (i.e.\ flux rope with lower minor radius), \citet{Regnault23} did not observe any particular rotation. Finally, the differences in solar wind configuration between the study of \citet{Regnault23} and the current simulations can explain why our observations on the rotation of the flux rope are different.

Several processes may have caused a rotation happening near the Sun. For example, \citet{Manchester17} suggested that the rotation can be due to the kink instability while for \citet{Shiota10} the smooth rotation of the structure is caused by the reconnection with the surrounding field. Interactions with the overlying magnetic field can also lead to a deflection of the CME into a streamer. Using a 2.5D MHD simulation of the corona on a specific date, \citet{Zuccarello11} observed a latitudinal migration of CMEs into a large helmet streamer due the imbalance in the magnetic pressure and tension force. However, such dynamics are not observed in any of our simulations despite the presence of important streamers in the maximum of activity configuration. Finally, further analysis is needed to determine whether the absence of rotation was anticipated or if it is a consequence of COCONUT being unable to accurately model the physical phenomena that could have caused it. 

The initial magnetic flux, and thus the initial speed, has limited influence on the density perturbation generated by the propagation of the CME. In both cases presented in Fig.~\ref{fig:density}, a sheath is formed ahead of the flux rope. However, the distribution of the radial velocity is very different (cf.\ Fig.~\ref{fig:density}, right panels). In the slower simulation, the maximum of velocity is mainly concentrated near the core of the flux rope. The transition from the solar wind speed to the flux rope velocity is smooth. Behind the flux rope, we can distinguish the region of the current sheet but there is no particularly high speed at the legs of the CME. This distribution of the speed is similar to the one observed by \citet{Maharana22} studying the evolution of a 3D flux rope model in EUHFORIA \citep[see also][for the same distribution with the spheromak CME model in the heliosphere]{Verbeke19}.

In the faster simulation (cf.\ Fig.~\ref{fig:density}, bottom right panel), the speed difference between the solar wind and the flux rope is much more pronounced. In this type of propagation, a piston-driven shock wave ahead of it is expected \citep{Chen11}. When comparing the right panels of the Fig.~\ref{fig:density}, we can see that the ellipsoidal area where the solar wind speed has been modified by the propagation of the flux rope is quite similar in both cases. However, the speed distribution inside this area is not the same. In the bottom panel (the faster CME), the maximum of speed is reached at two high speed flows starting from the sun rather than at the bulk of the flux rope. These regions cover the CME legs. The speed inside these streams is typically higher than the initial speed.

According to the "standard model" for CME, two other fast-mode shocks could be observed. These shocks occur when the upward reconnection outflow collides with the flux rope, and when the downward reconnection outflow collides with the post-flare loops. However, we suggest that the high-speed streams observed in our simulations are mainly numerical errors related to the difficulty encountered by the solver to handle high-speed gradients. Because of these poorly supported speed gradient, the solver produces regions where the pressure is numerically negative. As \citet{Regnault23} who faced similar problems in their fastest simulations, the pressure is set to $10^{-12}$ in normalised units when it should be negative. We also attempted to mitigate this behavior by increasing the spatial resolution with a 2.3 million-cell grid. However, even with this higher resolution, we still observed the presence of these high-speed streams. It is also important to note that \citet{Regnault23} encountered the same issue despite using an AMR grid, which offers better resolution at these high-speed stream locations. Finally, to limit the presence of these numerical artefacts, one solution could be to decrease the time step for better resolution, but at the expense of computing time.

In conclusion, while a too high initial velocity does not seem to greatly affect the distribution of the magnetic structure and density, it does generate non-physical dynamics. Therefore, future users should be aware of this and focus on flux ropes with initial speeds lower than 2000 km/s. High-speed gradients are only observed in the three fastest simulations ($\zeta=7$, $\zeta=8$, and $\zeta=9$). In all other simulations, regardless of the solar wind configuration, these high-speed streams are not observed. The speed distribution in these simulations is comparable to that of the top right panel in Fig. \ref{fig:density}. Finally, it is worth remembering that there are only very few CMEs that have a speed higher than $1500$ km/s \citep{Gopalswamy09}. COCONUT is therefore suitable for the study of the majority of cases observed.

\section{Thermodynamic and magnetic properties at 21.5 solar radii} \label{sec:profiles}
\subsection{Magnetic field components} \label{sec:components}

The simulations were conducted in order to demonstrate the potential use of COCONUT to provide heliospheric models with a realistic description of a CME. Some of these models, such as EUHFORIA, begins at 0.1~AU ($R=21.5\;R_{\odot}$). Therefore, in the following sections, the evolution of the plasma and the magnetic properties at a virtual satellite placed at $x=-21.5\;R_{\odot}$, $y=0\;R_{\odot}$, $z=0\;R_{\odot}$ will be detailed. This point is directly in the direction of the flux rope's propagation (cf.\ the white sphere in the Fig.~\ref{fig:density}).

\begin{figure}[h!]
\centering
\includegraphics[width=0.5\textwidth]{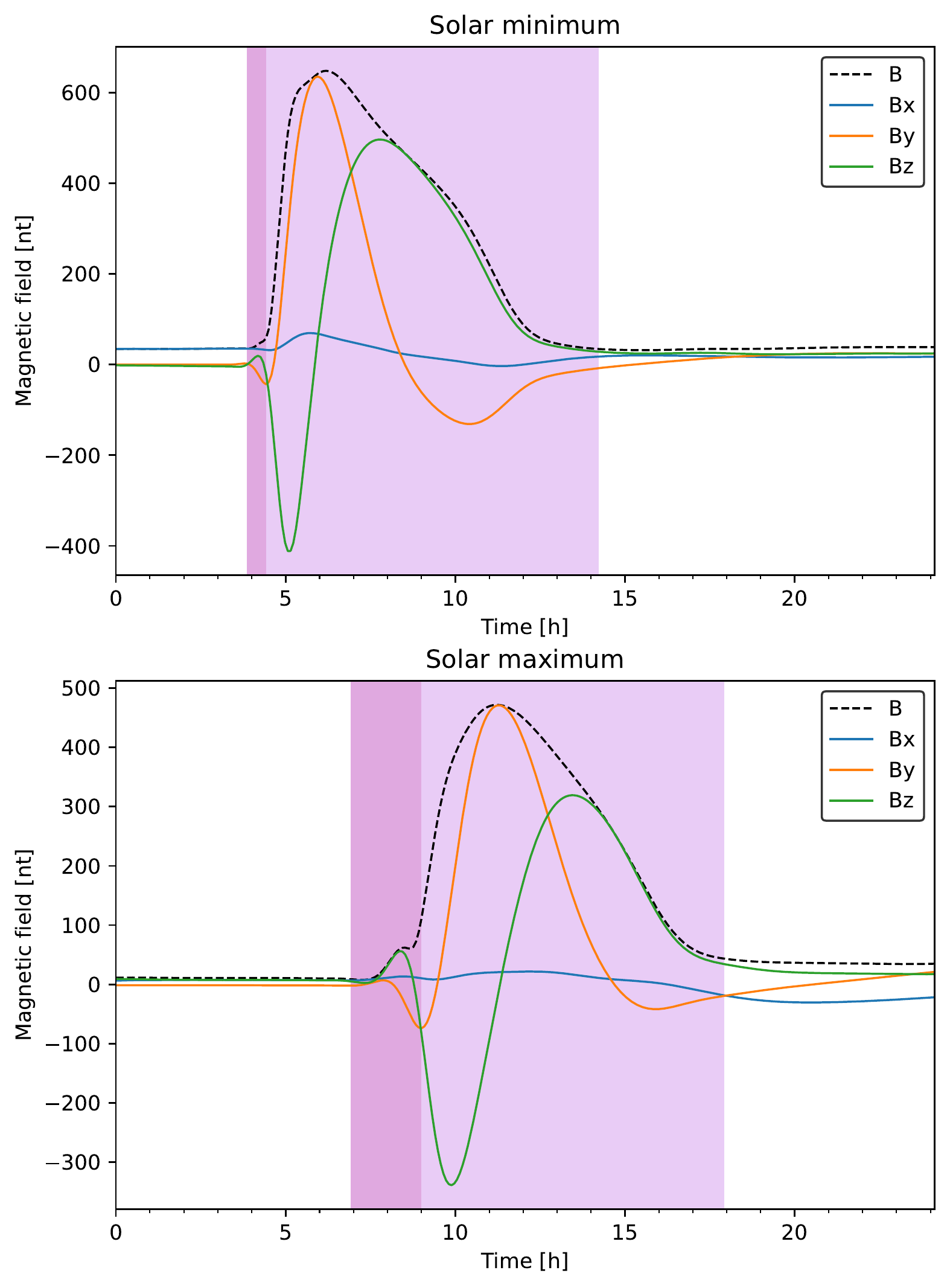}
\caption{Time evolution of the components of the magnetic field at 21.5 solar radii ($X=-21.5\;R_{\odot}$, $Y=0$, $Z=0$). The top panel corresponds to the flux rope with $\zeta=20$ in the solar wind background from a minimum of solar activity while the bottom panel is the case $\zeta=2$ in the maximum of activity. The dark purple band corresponds to the sheath region while the purple one is for the magnetic ejecta. The dynamics of the profiles are very similar in the two solar wind configurations : we observe a sheath followed by a magnetic ejecta composed of the flux rope initially inserted. }
\label{fig:B_comp}
\end{figure}

The first quantity of interest is the magnetic field. The Fig.~\ref{fig:B_comp} shows the evolution of the magnetic field components ($B_{x}$, $B_{y}$, $B_{z}$) as well as the magnetic field amplitude, $B$, for simulations in the minimum of activity (the top panel) and in the maximum activity (the bottom panel). The $x$-direction corresponds to the horizontal axis in the Fig.~\ref{fig:density}, the $y$-direction is the vertical axis and the $z$-direction is perpendicular to both the $x$- and $y$-directions. In the solar minimum, the flux rope is initially implemented with $\zeta=20$, and with $\zeta=2$ in the solar maximum. These simulations are used to illustrate all the cases as the dynamics described below can be found for all flux ropes. Only the amplitude of the different components changes between cases, not relative variations (cf.\ Sect.~\ref{sec:synthetic}).

The two selected flux ropes do not have the same intial velocity. The CME speed is $1220$ km/s for the flux rope evolving in the minimum activity while it is $455$ km/s for the CME in the maximum activity. Therefore, the CME in the minimum activity reaches the 0.1 ~AU boundary before the one in the maximum activity. This would be opposite if a faster CME had been taken in the maximum activity.

Regarding the profiles, the first thing to notice is that there is little difference between the simulation evolving in a solar minimum and that evolving in a solar maximum. This suggests that the differences in shapes and orientations observed in section \ref{sec:firststage} do not result in a significant change in the magnetic properties of the flux rope when crossing the distance of 21.5 solar radii.

In details, during the first hours, the magnetic field remains steady. It is equal to the field of the solar wind. Then, there is a fluctuation of the $B_{y}$ and $B_{z}$ components which leads to an increase of the magnetic field amplitude (cf.\ Fig.~\ref{fig:B_comp}). These fluctuations occur earlier in the activity minimum since the flux rope evolves faster. This period reflects the crossing of the sheath (see the dark purple band in the Fig.~\ref{fig:B_comp}). At around $t\sim9h$ in the maximum activity, the total magnetic field, $B$, begins to decrease before experiencing a second increase. As shown in the Fig.~4 in \citet{Regnault20}, the transition between two bumps indicates the arrival of the magnetic ejecta for relatively fast events with sheath. However, this pronounced change of slope is not observed in the minimum of activity (cf.\ top panel in Fig.~\ref{fig:B_comp}). The observed difference in the sheath between simulations evolving in minimum and maximum activity can be attributed to the difference in the global magnetic field configuration. The sheath results from the compression of magnetic field and plasma by the propagation of the flux rope. As the solar wind configurations are different, the magnetic field profiles in the sheath also differ.

The magnetic field profile within the magnetic ejecta is asymmetric. The maximum is reached in the first half corresponding to the front of the magnetic structure. Because of the speed difference, the maximum is reached more quickly in the simulation with $\zeta=20$ (cf.\ Fig.~\ref{fig:B_comp}, top panel). After the peak, in both cases, the magnetic field decreases before stabilizing. In the solar maximum, we note that the final value for the wake of the CME is higher than before its crossing as expected. Indeed, studying a set of more than 300 profiles observed at 1~AU, \citet{Regnault20} found that the most probable value for the total magnetic field is $4.9\pm0.6$ for the solar wind before the CME and $5.5\pm1.0$ after in the wake. This effect is enhanced for speed that are significantly faster than the solar wind before the CME.

Regarding the magnetic field components in the magnetic field (cf.\ Fig.~\ref{fig:B_comp}), there is only a minor variation in the $B_{x}$ component. According to the Lundquist model of flux rope, it should be zero while crossing the front part of the CME \citep{Lundquist51}. This suggests that the virtual satellite is crossing the flank of the CME or there may be a very slight rotation not captured by tracing the magnetic field lines. However, this contribution is negligible compared to the other components. The evolution of the magnetic field is primarily dominated by the $B_{y}$ and $B_{z}$ components.

In the magnetic ejecta (light purple shaded area), the $B_{y}$ starts by increasing to its maximum which coincides with the maximum of the total magnetic field. Then, the vertical component decreases until it reaches a negative value. In the final hours, $B_{y}$ increases to reach a slightly higher value than before the event. For $B_{z}$ (the green lines in the Fig.~\ref{fig:B_comp}), its dynamics is characterized by a change of sign. The evolution of $B_{z}$ is opposite to that of the $B_{y}$ component, beginning by decreasing before rising again once its minimum is reached. In both solar wind, the minimum and maximum peaks reached by $B_{z}$ are almost identical in absolute value. The profile of $B_{z}$ is asymmetric. Indeed, in the magnetic ejecta, $B_{z}$ is more often positive than negative. It is worth noting that the reverse of sign of the $B_{y}$ and $B_{z}$ profiles can be measured by in-situ satellite at 1~AU \citep[e.g.\ Fig.~1 in ][]{Regnault20}.

\begin{figure}[h!]
\centering
\includegraphics[width=0.5\textwidth]{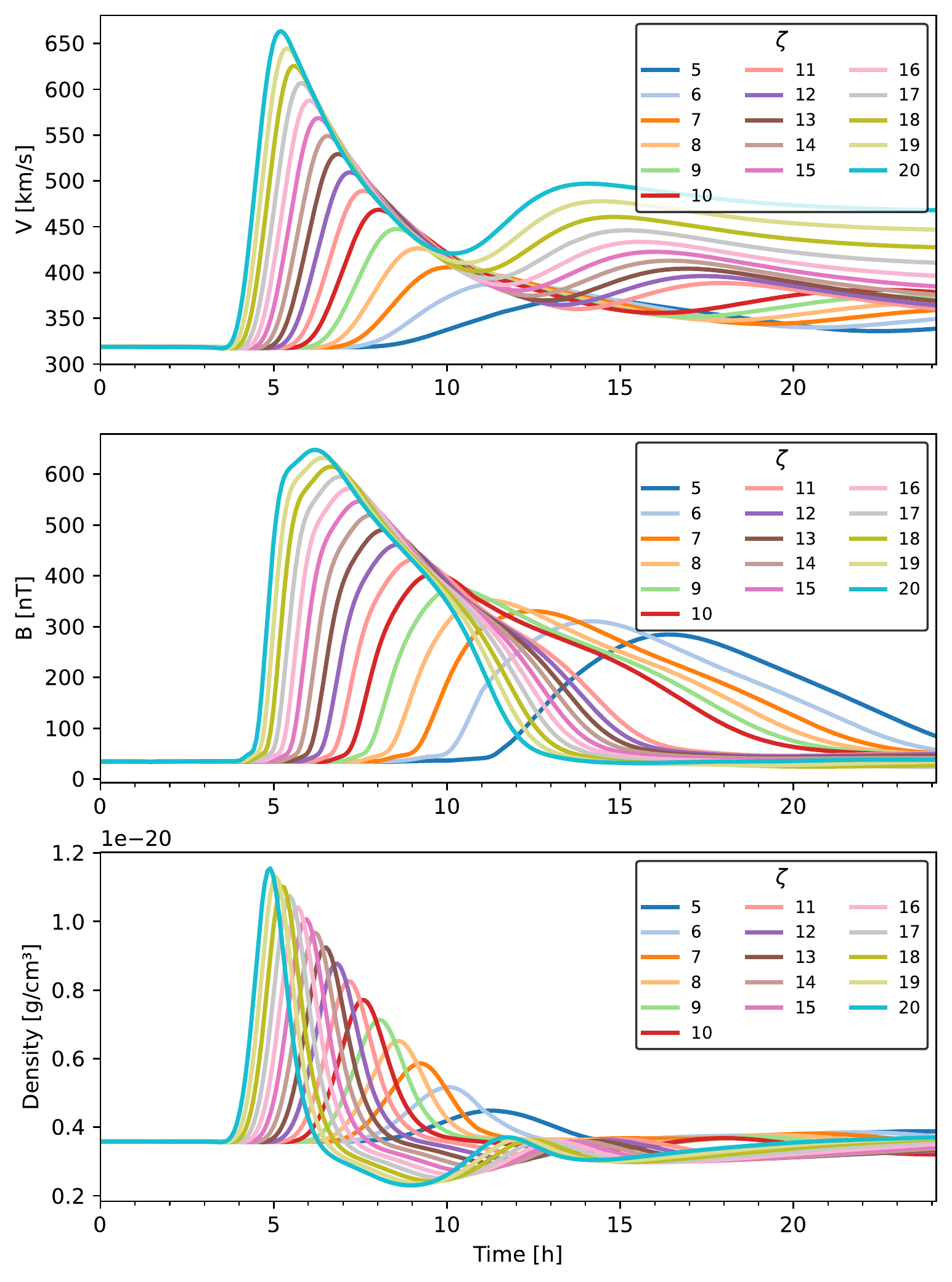}
\caption{Velocity, magnetic field and density evolutions as a function of time at a virtual satellite placed at $x\approx -21.5\;R_{\odot}$, $y\approx 0\;R_{\odot}$ and $z\approx 0\;R_{\odot}$. Each line corresponds to a different simulation in the background solar wind from a solar minimum. These profiles are consistent with the propagation of a flux rope with a sheath ahead of if.}
\label{fig:zeta_min}
\end{figure}

Finally, the variations of the magnetic components are in agreement with those expected by the TDm model \citep{Titov14}. This supports the conclusion that the magnetic structure is preserved during its expansion in the solar corona. The rearrangement of the field lines discussed in the section \ref{sec:firststage} does not imply a total destruction of the flux rope in both solar wind configurations.

\subsection{Density, magnetic field and velocity 1D profiles} \label{sec:synthetic}

\begin{figure}[ht!]
\centering
\includegraphics[width=0.5\textwidth]{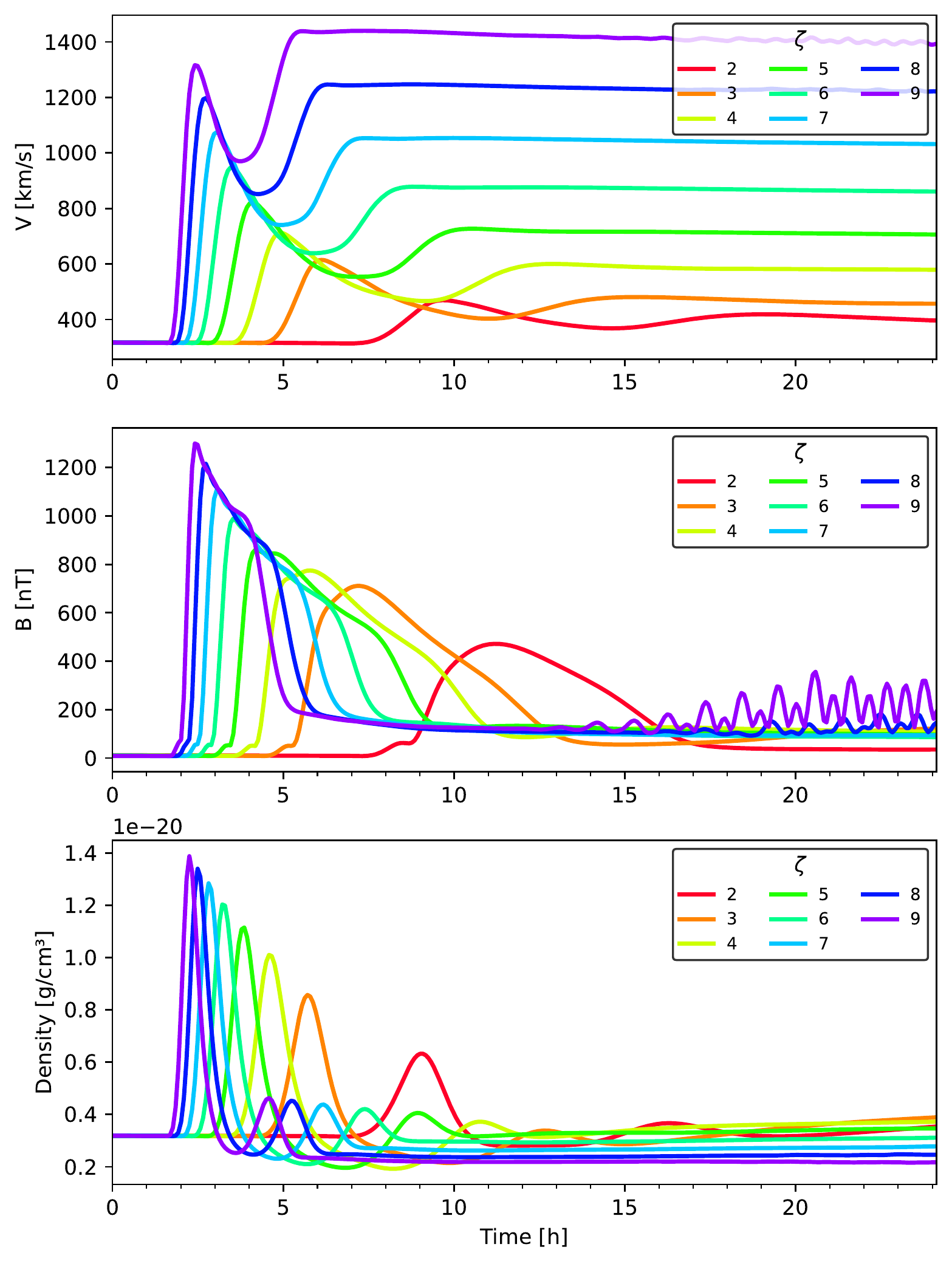}
\caption{Same as the Fig.~\ref{fig:zeta_min} but for the simulations in the maximum of activity. The profiles are similar to those obtained in the minimum activity meaning that COCONUT is effective in describing the propagation of the flux rope even during a solar maximum. }
\label{fig:zeta_max}
\end{figure}

In the previous section, we have seen that the magnetic ejecta that crosses the virtual satellite placed at $R=21.5\;R_{\odot}$ is indeed the flux rope that travelled from the solar surface. Now we will focus on the impact of the CME passage on the amplitude of the background magnetic field, velocity and density in all our simulations (cf.\ Figs.~\ref{fig:zeta_min} and \ref{fig:zeta_max}). The components of the magnetic field and the velocity, as well as the density are the quantities that are typically injected at the inner boundary of the heliospheric simulation to model a CME \citep[e.g][]{Verbeke19,Maharana22}.

Fig.~\ref{fig:zeta_min} shows the evolution of the three quantities in all the simulations for the solar minimum. Each color corresponds to flux ropes implemented with a different $\zeta$ parameter. First, it can be observed that the changes caused by the passage of the sheath and the flux rope do not start at the same time in all our cases. For the fastest CME, the arrival time is approximately $\sim3.8\;$h after the beginning of the simulation, while it is $\sim6\;$h for the slowest one.

The encounter with the sheath is characterized by an increase in density, magnetic field and velocity as expected by the in-situ measurement. The increase is more abrupt for simulations with a high $\zeta$ parameter, while it is smoother for the slowest cases. After a reaching a maximum whose value will be discussed in the next section (cf.\ Sect.~\ref{sec:max_ftsolarwind}), the three quantities decrease. However, the further evolution differs depending on the parameter.

As described in section \ref{sec:components}, the magnetic field, after it has reached its peak, decreases before stabilizing around a value slightly higher than the pre-event solar wind value. For the density (cf.\ Fig.~\ref{fig:zeta_min}, last panel), the simulations with the smaller $\zeta$, reach a maximum before returning to a value close to the one before the crossing. For the simulations with higher $\zeta$ (and therefore higher initial velocity), during a period of time that we associate with the crossing of the magnetic ejecta, the density is lower than before the arrival of the sheath. At the end of the simulations, the density increases slowly but does not return to its initial value within the next 10 hours. Studying the density profiles at 1~AU, \citet{Regnault20} found the same difference between simulations with relatively fast CMEs (compared to the speed of the solar wind) and slower CMEs. The distinctions observed in COCONUT are thus not surprising and are in agreement with the in-situ measurements further in the heliosphere. 

We also note that the density bump is thinner than the magnetic field bump. The density takes less time to return to values close to those of the solar wind than the magnetic field. This supports the conclusion that the disturbance encountered at $R=21.5\;R_{\odot}$ (0.1~AU) is indeed composed by a sheath where the matter has accumulated and a magnetic ejecta corresponding to the flux rope initially implemented.

The evolution of the velocity profile is divided in two bumps (cf.\ Fig.~\ref{fig:zeta_min}, top panel). The first one is the shock at the point of contact with the sheath, and the second is the arrival of the magnetic ejecta. The higher the speed of the shock is, the higher the speed in the wake of the CME. The maximum in the wake is approximately $70-75$ \% of the maximum speed of the shock. As for the density, for the relatively fast events, the speed of the solar wind post-CME is expected to be higher than before the CME \citep{Regnault20}.

The Fig.~\ref{fig:zeta_max} show the density, magnetic field and velocity 1D profiles for the 8 simulation performed in a solar maximum. Except for the values of the peaks reached, the profiles of the three quantities for the simulations in a maximum of activity show similar trends to those obtained in the minimum of activity. Most of the observations on profile variations are the same for both wind configurations, indicating that the difference in shape and orientation of the flux ropes (cf.\ Fig.~\ref{fig:comp_evol}) has little impact on the 1D profiles in the equatorial plane.

However, three differences can be observed between the simulation results in minimum and maximum activity. First, as previously mentioned in section \ref{sec:components}, there is a change of slope in the magnetic field increase that can be attributed to the transition between the sheath and the magnetic ejecta. This change is not observed in the simulations evolving in a minimum of activity. Secondly, the density profile has a second local maximum that is above the pre-event value. Finally the main difference concerns the evolution of the velocity. For the two fastest flux ropes ($\zeta=8$ and $\zeta=9$ in the Fig.~\ref{fig:zeta_max}), the speed in the wake of the CME is higher than during the crossing of the magnetic ejecta. This behavior was already observed in the 2D slices of the Fig.~\ref{fig:density}, where the maximum of speed is reached close to the legs of the flux and not ahead of it. Moreover, at the end of the fastest simulations, there are fluctuations of the magnetic field indicating that the solver has difficulty converging. The issue with convergence might be due to the fact that the modelled phenomena is inherently transient and thus, no immediate steady-state solution that the solver could converge to exists. In conclusion, this reinforces the idea that future studies should be limited to flux rope propagation with a velocity below $2000$ km/s, as suggested in section \ref{sec:limitation} in order to limit the unphysical behavior.

\subsection{Influence of the solar wind configuration} \label{sec:max_ftsolarwind}

\begin{table}[ht!]
\centering
\begin{tabular}{cccccc}
 \multicolumn{5}{c}{Quantities at $21.5\;R_{\odot}$} \\ \hline
$\zeta$ & {$V_{max}$ $[km/s]$} &  {$B_{max}$ $[nT]$} &  {$n_{max}$ $[10^{-21}g/cm^{3}]$} &  {$\Delta t_{sheath}$ $[h]$} \\ \hline
5 &  {371} &  {285} &  {4.5} &  {3.0} \\ 
6 &  {387} &  {310} &  {5.2} &  {2.6} \\ 
7 &  {406} &  {330} &  {5.9} &  {2.3} \\ 
8 &  {426} &  {352} &  {6.5} &  {2.1} \\ 
9 &  {447} &  {377} &  {7.1} &  {1.9} \\ 
10 &  {468} &  {404} &  {7.7} &  {1.8} \\ 
11 &  {489} &  {433} &  {8.3} &  {1.6} \\ 
12 &  {509} &  {462} &  {8.8} &  {1.6} \\ 
13 &  {529} &  {491} &  {9.2} &  {1.4} \\ 
14 &  {549} &  {520} &  {9.7} &  {1.4} \\ 
15 &  {569} &  {547} &  {10.0} &  {1.3} \\ 
16 &  {587} &  {572} &  {10.4} &  {1.3} \\
17 &  {607} &  {595} &  {10.7} &  {1.2} \\
18 &  {625} &  {615} &  {11.0} &  {1.2} \\
19 &  {645} &  {632} &  {11.3} &  {1.0} \\
20 &  {663} &  {647} &  {11.5} &  {1.0} \\ 
\end{tabular}
\caption{Summary table of the different simulations in the minimum activity. The different columns show the maximum of velocity, $V_{max}$, of magnetic field, $B_{max}$, and of density, $n_{max}$, as well as the the approximate size of the sheath $\Delta t_{sheath}$, reached at $R=21.5\;R_{\odot}$ in function of the parameter $\zeta$.}
\label{tab:minimum_21}
\end{table}

\begin{table}[ht!]
\centering
\begin{tabular}{cccccc}
 \multicolumn{5}{c}{Quantities at $21.5R_{\odot}$} \\ \hline 
 $\zeta$ &  {$V_{max}$ $[km/s]$} &  {$B_{max}$ $[nT]$} &  {$n_{max}$ $[10^{-21}g/cm^{3}]$} &  {$\Delta t_{sheath}$ $[h]$} \\ \hline
 2 &  {469} &  {472} &  {6.3} &  {1.8} \\ 
 3 &  {614} &  {710} &  {8.6} &  {1.1} \\ 
 4 &  {711} &  {775} &  {10.1} &  {0.9} \\ 
5 &  {822} &  {860} &  {11.1} &  {0.7} \\ 
6 &  {947} &  {989} &  {12.0} &  {0.7} \\ 
7 &  {1074} &  {1108} &  {12.8} &  {0.6} \\ 
8 &  {1197} &  {1215} &  {13.4} &  {0.5} \\ 
9 &  {1316} &  {1300} &  {13.9} &  {0.08} \\ 
\end{tabular}
\caption{Same as Fig.~\ref{tab:minimum_21} but for the simulations in the maximum activity.}
\label{tab:maximum_21}
\end{table}

\begin{figure}[h!]
\centering
\includegraphics[width=0.5\textwidth]{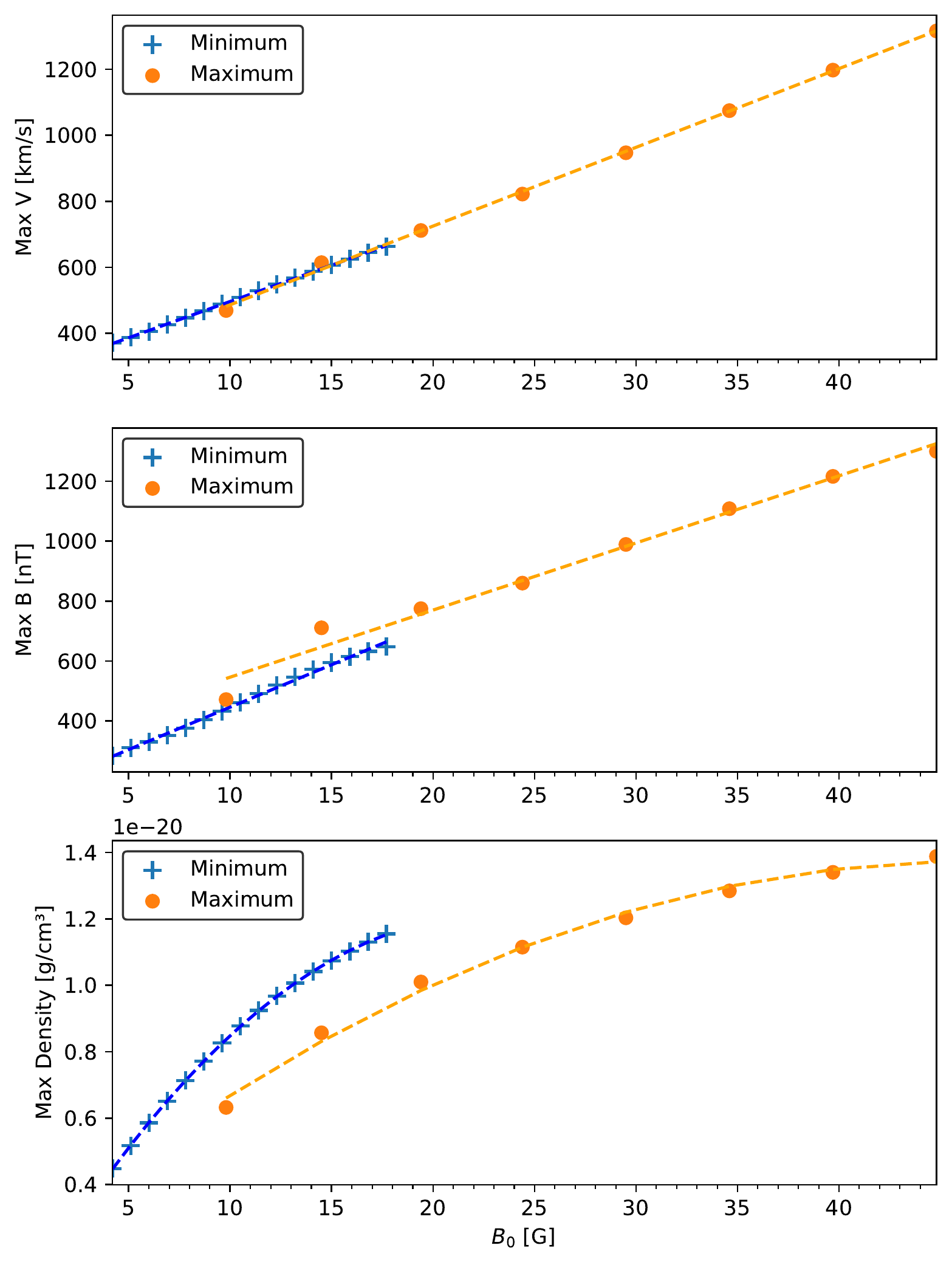}
\caption{Maximum values of the different temporal profiles presented in in Figs. \ref{fig:zeta_min} and \ref{fig:zeta_max}. From top to bottom : maximum of the velocity, of the magnetic field amplitude and of the density reached at $x\approx -21.5\;R_{\odot}$ as a function of the initial magnetic field of the flux rope. The "+" markers are for the flux rope in the minimum of activity while the orange dot correspond to the simulation in the maximum of activity. The dotted lines in orange and blue are polynomial regressions between the different values. The slope of the regressions depends on the solar wind configuration, meaning that the properties of the flux rope is related to its initial characteristics but also to the surrounding magnetic field.}
\label{fig:max_comp}
\end{figure}

In Figures \ref{fig:zeta_min} and \ref{fig:zeta_max}, we focused on the dynamics of 1D profiles of density, magnetic field and velocity at $R=21.5$ solar radii. The main conclusion was that the solar wind configuration has little influence on their evolution considering polytropic solar wind. Now we are going to have a closer look on the maximum values reached by these three quantities based on the initial parameters of the flux ropes and the solar wind configuration. The Fig.~\ref{fig:max_comp} shows the maxima of the magnetic field, velocity and density as a function of the initial magnetic field strength $B_{0}$ of the flux ropes in both solar wind backgrounds. Simulations from solar maximum are indicated by orange points while the blue crossed markers correspond to simulations in the solar minimum. The maximum values are also summarized in the tables \ref{tab:minimum_21} and \ref{tab:maximum_21}.

Our results show that, except for the two slowest flux ropes, the maximum speed is lower than the speed measured at the time t=$0.08$h. For example, in the case $\zeta=9$ at maximum of activity, the speed has decreased by approximately $55$ \%. The fact that two simulations have a higher speed at 21.5 solar radii than near the Sun indicates that the flux ropes accelerate during a given period. This acceleration could be caused by the momentum of the reconnection outflow \citep{Shibata01} or by other factors such as ideal MHD instabilities \citep{Fan07}.

Our findings are consistent with the theory that suggests two forces act on a CME \citep{Sachdeva17}. The first force is the Lorentz force, which allows for the radial expansion of the CME and whose profile peaks quickly and then declines. The second force is a drag force due to the surrounding solar wind, which tends to hold down the expanding flux rope. This force explains why CMEs gradually slow down during their propagation. Over the years, the characteristics of these two forces have been studied analytically \citep[e.g.][]{Chen10,Subramanian12} and using observational data \citep[e.g.][]{Sachdeva15,Sachdeva17}.

In Fig.~\ref{fig:max_comp}, top panel, we see that the maximum of speed $V_{max}$ appears to increase linearly with the initial magnetic strength. To confirm this, we performed linear regressions in both solar wind configurations. In the minimum of activity the formula obtained is $V_{max}\approx21.9 \times B_{0}+277.1$ with a coefficient of determination $r^{2}=0.99$. In the maximum of activity the relation is $V_{max}\approx23.8 \times B_{0}+248.3$ with a coefficient of determination of $0.99$. For low $B_{0}$ (lower than $15.16$ G), the CMEs in the minimum of activity would be faster than CME in the maximum of activity. On the contrary, for large $B_{0}$ values (higher than $15.16$ G), the flux ropes would be faster in the solar maximum. This is more consistent with the observations. Indeed \citet{Regnault20}, from in-situ measurements of the ACE satellite at the L1 Lagrangian point found that CMEs during active periods tend to be faster than during quiet periods \citep[cf.\ also][]{Hundhausen99}.

The maximum of the magnetic field also appears to increase linearly with the initial field (cf.\ Fig.~\ref{fig:max_comp}, middle panel). The correlation is $B_{max}\approx28.3 \times B_{0}+163.4$ with $r^{2}=0.99$ in the solar minimum and $B_{max} \approx 22.3 \times B_{0}+322.5$ with $r^{2}=0.98$ in the solar maximum. Similar to the speed, the higher the magnetic flux of the flux rope is, the higher the magnetic field will be at 21.5 solar radii. Comparing the initial magnetic field value to the one measured at the virtual satellite, we observe a decrease in the magnetic field during the propagation (cf.\ Tabs. \ref{tab:minimum_21} and \ref{tab:maximum_21}). Different authors such as \citet{Leitner07,Liu05,Winslow15} have derived power laws in order to determine the decrease in magnetic field during the propagation phase. However these power laws use measurements made at least at 0.3~AU away from the Sun while our simulation grids stops at $25\;R_{\odot}$ ($\approx 0.12\;$AU). Thus, the comparison between our simulation and power laws deduced from in-situ observation is limited. However the latter turns out to be working well forecasting the magnetic field at 21.5 solar radii. \citet{Leitner07,Liu05,Winslow15} suggested that the magnetic field amplitude of the CMEs is between $10^{3}$ and $10^{4}$ nanotesla at 0.1~AU, which is consistent with the maximum encountered in our set of simulations.

 Unlike the other two quantities, the evolution of the maximum density can not be fitted by a linear regression (cf.\ Fig.~\ref{fig:max_comp}, bottom panel). This is simply due to the fact that no material is added during the implementation of the flux rope. This results in an increase in density at the crossing point that is entirely related to the accumulation of material in the sheath created by the propagation of the flux rope. With the mass being conserved in the domain, it cannot have an infinite increase. By fitting a polynomial regressions in both solar wind configurations, we see a horizontal asymptote that seems to be around $1.4-1.5e^{-20} g/cm^{3}$. Moreover, we observe that for a flux rope with the same magnetic field amplitude, the maximum of density reached is higher in the solar minimum than in the maximum activity. This is due to the fact that after the relaxations runs, the density for the minimum activity solar wind is slightly higher than for the maximum activity solar wind. The density distribution being not the same between the two configurations, the accumulation in the sheath is also different.
 
 \begin{figure}[h!]
\centering
\includegraphics[width=0.5\textwidth]{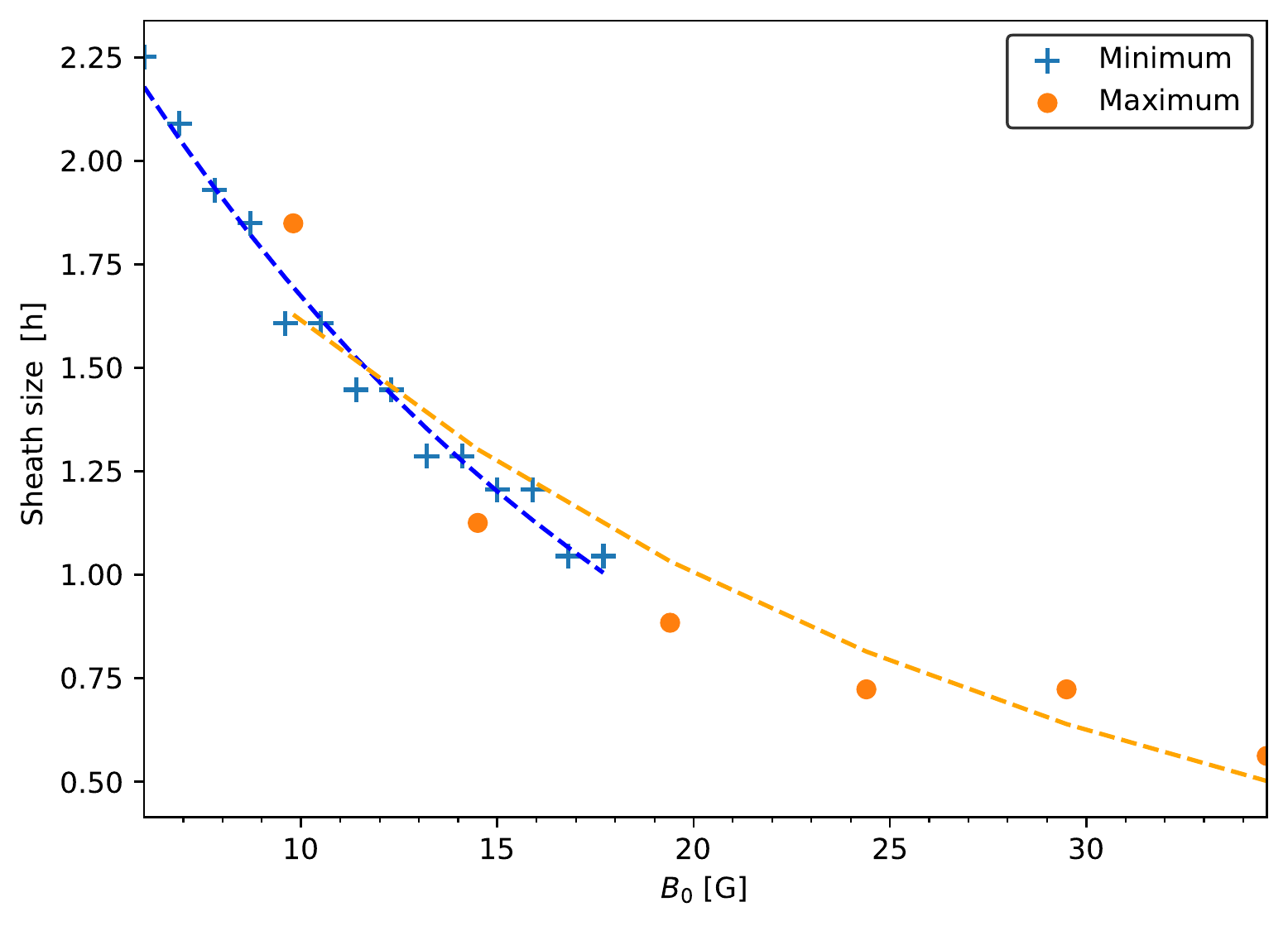}
\caption{Approximate size of the sheath (in hours) as a function of the initial magnetic field. The orange dots indicate the simulations at maximum of activity while the blue markers corresponds to simulations at minimum of activity. The orange and blue lines are exponential curves fitting the data. As expected, the width of the sheath decreases with increasing initial speed.}
\label{fig:size}
\end{figure}

 The faster the CMEs propagate, the higher the value of the density. But we also see that the full width at half maximum of density decreases as the initial magnetic flux increases (cf.\ Figs. \ref{fig:zeta_min} and \ref{fig:zeta_max}) as expected from some analytical sheath model such as \citet{Takahashi17}. In order to quantify the compression, we derived an approximate size of the sheath by measuring the time between the beginning of the increase of the magnetized field and the first local minimum of the $B_{y}$ component (cf.\ Fig.~\ref{fig:B_comp}). The results are summarized in the tables \ref{tab:minimum_21} 
 and \ref{tab:maximum_21}, and in the Fig.~\ref{fig:size}. We find that the sheath size roughly fits a power law with a slope dependent on the solar wind configurations. 

To summarize, the properties of the sheath and the flux rope seem directly related to the initial parameters in both solar wind configurations. However, the distribution of the surrounding magnetic field and density directly influences the value of the magnetic field, the speed and the density of the flux ropes as expected from the observations. 
 
\section{Summary and Conclusions}

The main objective of the present study was to demonstrate how the new 3D MHD coronal model COCONUT can be used to track the evolution of flux ropes in the corona. We first introduced the time-dependent implicit COCONUT solver (cf.\ Sect.~\ref{sec:Coconut}). Before implementing a CME, it was necessary to create the surrounding magnetic field in which the flux rope will evolve. We have thus run COCONUT in its relaxation mode to obtain the solar wind from photospheric magnetograms for two opposing solar wind configurations: one corresponding to a solar minimum ($2^{\rm nd}$ July 2019, CR2219), and the other representing a solar maximum ($20^{\rm th}$ of March 2015, CR2161). In both cases, we used pre-processed HMI magnetograms as initial boundary conditions. We also verified that the configurations created were in agreement with white light images and tomography measurements (cf.\ Sect.~\ref{sec:validation}). 

Once the background MHD corona was created, we could implement flux ropes using the method described in section~\ref{sec:implementation}. The chosen model for the flux rope was the Titov-D\'emoulin model \citep{Titov14}. We performed 24 different simulations in total with 16 in the minimum solar activity corona and 8 in the maximum activity corona. These simulations varied only in the initial magnetic flux of the implemented flux ropes. In all the cases, the propagation of the flux rope was driven by an imbalance caused by the Lorentz force.

In this study, we tracked the propagation of magnetic field lines by taking snapshots of the magnetic field lines at different instants of our simulations (cf.\ Sect.~\ref{sec:Prop}). A virtual satellite was also placed at $R=21.5\;$solar radii to extract 1D time profiles of density, magnetic field and velocity as the structures crossed this specific point in the domain (cf.\ Sect.~\ref{sec:profiles}). Finally, the main results of this study can be summarised as follows : 
\begin{itemize}
  \item During the first stages of the evolution, the shape of the flux rope is strongly affected by the surrounding field. In the minimum activity configuration, the flux rope has a symmetrical propagation unlike cases evolving in the maximum activity solar wind. However, in both cases, we observe dynamics consistent with the standard eruption model, such as the development of a current sheet layer below the core of the CME. This region is conductive to magnetic reconnection (purely due to numerical dissipation in our cases), which is observed numerically by the presence of post-flare loops in the simulation domain. 
  \item Further out in the corona, the flux rope continues to expand radially without any particular rotation and develops a sheath ahead of it, reflecting the accumulation of matter upstream of a piston-driven shock.
  \item In 2D slices of the radial speed, we find that artificial high-speed streams develop in the wake of the CME for the fastest flux ropes leading to convergence problems. We suggest limiting the future studies to flux rope initially propagating at speed lower than $2000\;$km/s, which correspond to most of the events observed.
  \item Despite the differences in geometry, the evolution of the magnetic field components at 0.1~AU is similar in both solar wind configurations. They indicate the presence of a sheath followed by a magnetic ejecta composed of the initially implemented flux rope. The values of the solar wind are impacted by the wake of the CME after it passed by.
  \item The 1D time profiles of the density, the magnetic field and the velocity obtained at 0.1~AU match related observations and in-situ measurement further in the heliosphere. Only local peaks and the duration of the disturbances differ between all simulations.
  \item By comparing the maximum of density and of magnetic field strength as a function of the initial magnetic field strength in both solar winds, we find simple linear relationships between quantities. The greater the initial imbalance of the Lorentz forces is, the higher the magnetic field strength will be at 0.1~AU. These relationships also suggest that CME speeds are higher in the maximum activity than in the minimum activity case when the initial magnetic field strength is higher than $15.16\;$G. Due to the difference in the density distribution, the cases evolving in the minimum activity have a higher peak of density than those in the maximum activity.
  \item As expected, the size of the sheath depends on the initial velocity of the flux rope. The fastest CMEs result in thinner sheath compared to the slowest cases.
\end{itemize}

To conclude, COCONUT is an efficient solver for studying the propagation of a flux rope in a realistic solar corona/wind configuration. Being faster than a state-of-the-art explicit solver, COCONUT appears as a valuable tool for understanding the physical mechanisms that occur during the initial CME propagation and for obtaining a detailed representation of a numerical CME at 0.1~AU after its interaction with the background solar corona reconstructed from observed magnetograms.

Using this work as basis, a logical next step could be to model a truly observed event. To do this, it would then be necessary to align the feet of the flux rope with the polarities of the active region. The following steps would then consist in ensuring that the properties of the flux rope (its shape, intensity, etc.) but also the properties of the simulated active region (its position, size, magnetic flux) coincide with what is measured at the photosphere but also observed in EUV in the low corona. It would also be necessary to modify the boundary conditions so that the polarities evolve similarly to what is observed at the surface of the Sun. Additionally, our analysis of the impact of magnetic flux on the properties of the CME at $21.5\;$solar radii would enable a more precise selection of the flux rope initial parameters to match the characteristics of the real event. However, the analysis should be expanded to include also other parameters such as the size and tilt of the flux rope.

It should also be noted that, like the PLUTO solver \citep{Regnault23}, COCONUT is currently unable to perform the relaxation of the flux rope before provoking its eruption. At this time, the super-imposed flux rope is out of equilibrium from the start. This is partly due to fixed inner boundary conditions which do not allow considering the emergence and shearing of the magnetic field at the photosphere, and also the need to extend COCONUT to model the solar plasma below the corona, i.e., to include the transition region and chromosphere.

To achieve better results, particularly in future work, the solver is currently being improved by adding more physical terms related to the heat transfer such as anisotropic thermal conduction, coronal heating and radiation. Indeed, the current version of COCONUT includes only polytropic MHD which leads to a biased distribution of density in the domain. In parallel, work is also underway to improve the balance of electromagnetic and gravitational forces in the domain.  Combined, these adjustments would allow for a more realistic distribution of the magnetic field and density, as well as a bimodal distribution for the solar wind. This would lead to a better overall description of CME propagation. Another notable addition would be to consider a non-zero resistivity in the domain to obtain a better description of magnetic reconnection, which is currently entirely due to numerical dissipation. Remark, however, that the resistivity in the hot solar corona is extremely low and current CPU power does not make it possible to increase the grid resolution so much that the numerical resistivity is even lower.

COCONUT was originally developed to replace the simple empirical coronal model used by the space weather simulation model EUHFORIA \citep{Pomoell18}. With improved initial conditions for the solar wind, EUHFORIA should yield more accurate predictions. By using our work as a foundation, we can also use the COCONUT to model the evolution of flux rope CMEs closer to the Sun and then insert it into the inner boundary of EUHFORIA. However, two limitations of COCONUT must be taken into account to facilitate such a coupling of the models. First, the grid resolution of COCONUT decreases as one moves away from the Sun. Thus, at 0.1~AU, the spatial resolution is lower than the standard EUHFORIA resolution at the inner boundary of the heliospheric wind and CME evolution model. A spatial interpolation that could lead to information loss is therefore necessary for the coupling. To remedy this, an increase in resolution of the COCONUT grid can be considered, as well as the use of more advanced grid refinement techniques. This could also mitigate the presence of the high-speed streams mentioned earlier.

The second limitation is the computation time. To obtain our results, it took 2 hours to create the background corona and between 20-26 hours for the propagation of the flux ropes. With a total time of 22-28 hours, COCONUT is a good first step toward a fully operational space weather asset, since CMEs typically take more than 2 days to reach the Earth. It is worth noting that COCONUT would be much faster than current state-of-the-art coronal models based on explicit solvers. Indeed, \citet{Perri23} showed that COCONUT (with an implicit solver) is up to $35\times$ faster than PLUTO (using an explicit solver) to model a realistic solar wind configuration. Currently, the performance of the two solvers with the same initial set-up was only made for the reconstruction of the background corona/solar wind (i.e., using the steady-state (implicit version of COCONUT). However, according to the results presented here, it is expected that the time-accurate version of COCONUT (with a flux rope inserted) would also be faster than PLUTO for studying the propagation of a CME.

In the future, the simulation model can be made even more efficient as there are several ways to further reduce the computation time. The first one is to increase the resources dedicated to the code execution (e.g., the number of cores) as COOLFluiD scales good on parallel architectures. Additionally, it is possible to optimize the code parameters, such as the time step for the study of a particular event. In our work, we chose the same parameters for all simulations, but a large fraction of the flux ropes (the fastest ones) reach the boundary well before 20 hours of computation. Moreover, a more detailed study of the impact of the time step combined with the grid resolution needs to be carried out to optimize CPU consumption.

Once the mentioned limitations are addressed and COCONUT is coupled with EUHFORIA, the effectiveness of the latter in predicting CME geoeffectiveness would be improved. Two approaches can be considered for inserting the flux rope that has evolved in COCONUT into the heliospheric wind and CME evolution module of EUHFORIA. The first approach is to extract at 0.1~AU only the properties of the flux rope and then to implement them into EUHFORIA, similar to how it is currently done for the spheromak CME model. However, this requires the ability to identify and isolate the CME structure. The second approach is to include the entire solar wind with the flux rope, i.e., to evolve the entire inlet boundary of EUHFORIA heliosphere (at 0.1~AU) in time using COCONUT. In both cases, the inserted flux rope has interacted with the solar wind before reaching 0.1~AU and should, therefore, have more realistic magnetic and geometric properties than the current self-similar models that are inserted at 0.1~AU without taking into account the initial evolution and interaction with the background corona. Overall, if coupled with an heliospheric model, COCONUT can be a valuable tool for space weather forecasting.

\begin{acknowledgements}
The authors thank the anonymous referee for their valuable comments. The authors acknowledges support from the European Union’s Horizon 2020 research and innovation programme under grant agreement N$^o$ 870405 (EUHFORIA~2.0). This work has been granted by the AFOSR basic research initiative project FA9550-18-1-0093. These results were also obtained in the framework of the projects
C14/19/089 (C1 project Internal Funds KU Leuven), G.0B58.23N (FWO-Vlaanderen), SIDC Data Exploitation (ESA Prodex-12), and Belspo project B2/191/P1/SWiM.
F.~R acknowledge grants 80NSSC20K0431, 80NSSC21K0463 and 80NSSC20K0700.
The resources and services used in this work were provided by the VSC (Flemish Supercomputer Centre), funded by the Research Foundation - Flanders (FWO) and the Flemish Government.
HMI data are courtesy of the Joint Science Operations Center (JSOC) Science Data Processing team at Stanford University.
We thank H. Morgan for the tomography data used. 
\end{acknowledgements}

%
%


\bibliographystyle{aa} 
\bibliography{biblio.bib}


\end{document}